\documentclass{aa}  
\usepackage{graphicx}
\usepackage{txfonts}
\usepackage{natbib}
\usepackage{graphicx}
\usepackage{hyperref}
\newcommand{\Ni}{{$^{56}$Ni}}

\usepackage{color}

\begin{document}

        \title{Mass discrepancy analysis for a select sample of \\ 
                   Type II-Plateau supernovae}
        
        \author{Laureano Martinez \inst{1,2}
                        \&
                        Melina C. Bersten \inst{1,2,3}
                        }

        \institute{Facultad de Ciencias Astronómicas y Geofísicas,
                        Universidad Nacional de La Plata,
                        Paseo del Bosque S/N, B1900FWA, La Plata, Argentina\\
                        \email{laureano@carina.fcaglp.unlp.edu.ar}
                \and
                        Instituto de Astrofísica de La Plata (IALP), CCT-CONICET-UNLP. 
                        Paseo del Bosque S/N, B1900FWA, La Plata, Argentina
                \and
                        Kavli Institute for the Physics and Mathematics of the Universe (WPI),
                        The University of Tokyo, 5-1-5 Kashiwanoha, Kashiwa, Chiba 277-8583, Japan\\
                        \email{mbersten@fcaglp.unlp.edu.ar}
                        }
%

\titlerunning{Mass discrepancy analysis of SNe II-Plateau}
\authorrunning{Martinez \& Bersten}

\date{Received XXX; accepted XXX}
 


\abstract 
{The detailed study of supernovae (SNe) and their progenitors allows a better understanding of the evolution of massive stars and how these end their lives. Despite its importance, the range of physical parameters for the most common type of explosion, the type II supernovae (SNe II), is still unknown. In particular, previous studies of type II-Plateau supernovae (SNe II-P) showed a discrepancy between the progenitor masses inferred from hydrodynamic models and those determined from the analysis of direct detections in archival images.

Our goal is to derive physical parameters (progenitor mass, radius, explosion energy and total mass of nickel) through hydrodynamical modelling of light curves and expansion velocity evolution for a select group of six SNe II-P (SN\,2004A, SN\,2004et, SN\,2005cs, SN\,2008bk, SN\,2012aw, and SN\,2012ec) that fulfilled the following three criteria: 1) enough photometric and spectroscopic monitoring is available to allow for a reliable hydrodynamical modelling; 2)  a direct progenitor detection has been achieved; and 3) there exists confirmation of the progenitor identification via its disappearance in post-explosion images. We then compare the masses obtained by our hydrodynamic models with those obtained by direct detections of the progenitors to test the existence of such a discrepancy. As opposed to some previous works, we find good agreement between both methods.

We obtain a wide range in the physical parameters for our SN sample. We infer presupernova masses between 10 and 23\,$M_{\odot}$, progenitor radii between 400 and 1250\,$R_{\odot}$, explosion energies between 0.2 and 1.4\,foe, and $^{56}$Ni masses between 0.0015 and 0.085\,$M_{\odot}$. An analysis of possible correlations between different explosion parameters is presented. The clearest relation found is that between the mass and the explosion energy, in the sense that more-massive objects produce higher-energy explosions, in agreement with previous studies. Finally, we also compare our results with previous physical--observed parameter relations widely used in the literature. We find significant differences between both methods, which indicates that caution should be exercised when using these relations.}


\maketitle
\section{Introduction} 
\label{sec:intro} 

Type II supernovae (SNe II) are the most common type of SNe found in nature \citep{arcavi10,li11}. They are observationally classified according to their spectral characteristics, showing strong and prominent P-Cygni hydrogen lines \citep{filippenko97}. Historically, SNe II have been subclassified according to the shapes of their light curves (LCs) into: II-Plateau (SNe II-P), characterised by a `plateau' in the optical light curve where the luminosity remains nearly constant for a period of $\sim$100 days, and II-Linear, showing linearly declining light curves \citep{barbon79}. However, recent studies have questioned this subdivision and propose the existence of a continuous sequence of these objects \citep{anderson14}. 

The LC morphologies of the SNe II-P are easy to reproduce assuming a red supergiant (RSG) progenitor with an extensive hydrogen envelope. This has been shown, some time ago,  by hydrodynamical models \citep{grassberg71,falk77} and confirmed more recently by direct detection of the progenitor star \citep{vandyk03,smartt04}. However, the main factor to change the slope during the plateau phase has not been clearly identified, although some ideas have been proposed such as, for example, the nickel mixing \citep[see][]{bersten11,kozyreva18}.

Additional interest has arisen in SNe II-P since they have been proposed as good distance indicators with potential application to cosmology, in an alternative way to the best-known method that involves Type Ia SNe. Several methods have been tested to obtain an accurate measurement of the distance \citep{hamuy02,rodriguez14,dejaeger15}. 

One fundamental question in astrophysics that remains unanswered is which evolutionary processes of massive stars determine the type of SN that they produce. It is currently believed that hydrogen-rich type II SNe are produced by the least massive stars among those suffering gravitational collapse, and also that these stars have been able to hold a significant fraction of their hydrogen-rich envelopes during evolution. In contrast, in Type Ib and Ic SNe (SNe Ib/c) progenitors have completely lost their hydrogen-rich envelopes. It is known that mass loss is more intense the more massive the star is, although there are other factors such as rotation or metallicity that can also affect mass loss. This is why traditionally it has been thought that SNe Ib/c should come from more massive stars than SNe II. Currently this vision is changing, since there is increasing evidence that this kind of object may come from binary evolution. In close binary systems, stars are expected to exchange mass, providing an efficient mechanism to allow for the removal of their outer layers regardless of the specific 
mass  of the stars. Recent studies in open clusters have shown that the fraction of interacting binary systems can reach 70\% of the total \citep{sana12}. Also, all studies that derived masses of SNe Ib/c from LCs modelling find low masses before the explosion which is interpreted as objects coming from binary systems \citep[see, e.g.][]{lyman16,taddia18}.

Our knowledge of the physical properties of SNe II is not entirely satisfactory. There are important discrepancies in the literature regarding masses and radii of the progenitors, depending on the different methods used for the analysis. The most immediate method to determine what type of star gives rise to a certain SN is to detect the progenitor star at the explosion site using archival images \citep{vandyk03,smartt04,maund14a}. Recent detections of progentitors of several SNe II-P have confirmed that, in effect, they come from RSG stars, although their masses in the zero age main sequence (ZAMS) have been estimated with values lower than 17\,$M_{\odot}$ \citep{smartt09,smartt15}, unlike the cutoff of 25$-$30\,$M_{\odot}$ predicted by evolutionary models. \cite{davies18} investigated one particular source of systematic error present in converting pre-explosion photometry into an initial mass, namely that of the bolometric correction used to convert single-band flux into bolometric luminosity. They show that the updated initial mass function results in an increased upper mass cutoff of 19\,$M_{\odot}$. Detection of the progenitor in archival images requires the acquisition of later images of the SN, which are necessary to confirm the identification of the progenitor when observing its disappearance. However, this method can only be applied when pre-explosion images are available and with nearby SNe (d $\lesssim$ 30 Mpc) because of the lack of resolution and sensitivity for more distant sources.

Nebular-phase spectral modelling can also be used to constrain the progenitor masses of SNe II. In the nebular phase, the inner ejecta become visible and the nucleosynthesis yields can be analysed. Thus, using the dependency of oxygen production on progenitor ZAMS mass, it is possible to distinguish between different progenitors \citep[see, e.g.][]{jerkstrand12,jerkstrand14}.

Finally, hydrodynamic modelling of LCs is one of the most commonly used indirect methods to derive physical properties. The LCs of SNe are extremely sensitive to the physical properties of their progenitors (masses and radii), as well as the properties of the explosion itself \cite[released energy, amount of synthesised radioactive nickel, and its distribution, see e.g.][among others]{shigeyama90,bersten12}. A problem that has been noted in the literature \citep[][among others]{utrobin08,utrobin09,maguire10} is that the mass estimated by hydrodynamical models is usually larger than the estimate or upper limit given by pre-SN imaging.

In this work, we are mainly interested in analysing the discrepancy suggested in the literature surrounding the progenitor mass. To do this we select a group of SNe II-P for which their exists the most information and derive physical parameters through the hydrodynamic modelling of their LCs to compare these results with those obtained from pre-explosion information available in the literature. This analysis is the first step that we propose before going on to analyse a large sample of H-rich SNe. A preliminary analysis of this study can be seen in \citet{martinez18} for a subgroup of objects.

The paper is organised as follows. In the following section we present our sample of SNe with its selection criteria and provide a brief description of each one. In Section \ref{sec:models}, we describe our hydrodynamic code and the pre-SN models used. In Section \ref{sec:results}, we present the main results of this paper. Section \ref{sec:analysis} summarises the analysis performed on our results, and in Section \ref{sec:conclusion}, we provide some concluding remarks.

\section{Supernova sample}
\label{sec:sample}

There are hundreds of SNe II-P, however only a few of them are useful for our purpose. We first chose a group of objects that followed our selection criteria: SNe II-P for which there exists (\textit{i}) enough available photometric and spectroscopic monitoring during the plateau and the radioactive phase to allow reliable hydrodynamical modelling of their LCs and photospheric velocity evolution; (\textit{ii}) pre-explosion images with direct information from the putative progenitor star; (\textit{iii}) post-explosion images confirming the disappearance of the progenitor. After an exhaustive search in the literature, we found six SNe II-P that match our criteria: SN 2004A, SN 2004et, SN 2005cs, SN 2008bk, SN 2012aw, and SN 2012ec.

Photometric data and expansion velocities for these objects were obtained from: \cite{gurugubelli08} for SN 2004A, \cite{maguire10} for SN 2004et, \cite{pastorello06,pastorello09} for SN 2005cs, the CHilean Automatic Supernova sEarch (CHASE) project \citep[][G. Pignata private communication]{chase} for SN 2008bk, \cite{bose13} and \cite{dallora14} for SN 2012aw, and \cite{barbarino15} for SN 2012ec.

From the literature we also obtained distances, galactic and host-galaxy extinctions, and an estimate of the explosion epochs, which are presented in Table \ref{table:info}. When different values of distance for the same SN coexist in the literature, we adopt the value that was determined by more distance estimation methods. If only one value is provided, we adopt the value used by the author of the paper from which we extracted photometry and expansion velocities.

\begin{table*}
        \caption{Galactic and host galaxy extinctions, distances, and explosion times for the sample.}             
        \label{table:info}      
        \centering          
        \begin{tabular}{c c c c c c c c c}
        \hline\hline\noalign{\smallskip}
        SN & Host galaxy & Distance & E(B$-$V)\,$_{\rm gal}$ & E(B$-$V)\,$_{\rm host}$ & A$_{V}^{{\rm gal}}$ & A$_{V}^{{\rm host}}$ & Explosion epoch & References\\
        & & [Mpc]       &       &       &       &       & [JD-2450000] & \\
        \hline\noalign{\smallskip}
        2004A  & NGC 6207 & 20.3 $\pm$ 3.4 & --- & --- & 0.042 & 0.598 & 3006 & 1, 2, 3 \\
        2004et & NGC 6946 & 5.6 & 0.34 & 0.07 & --- & --- & 3270.5 & 4, 5, 6 \\
        2005cs & M 51     & 7.1 $\pm$ 1.2 & 0.035 & 0.015 & --- & --- & 3549 & 7 \\
        2008bk & NGC 7793 & 3.43                 & --- & --- & 0.065 & 0.0 & 4543 & 1, 8 \\
        2012aw & NGC 3351 & 9.9 $\pm$ 0.1 & 0.0278$\pm$ 0.0002 & 0.046 $\pm$ 0.008 & --- & --- & 6002.5 & 9 \\
        2012ec & NGC 1084 & 17.29 & 0.024 & 0.12$^{+0.15}_{-0.12}$ & 0.074 & 0.372 & 6147.5 & 1, 10, 11 \\
        \noalign{\smallskip}
        \hline
        \end{tabular}
\tablebib{(1) This work; (2) \cite{hendry06}; (3) \cite{maund17};
(4) \cite{zwitter04}; (5) \cite{li05}; (6) \cite{sahu06}; \\
(7) \cite{pastorello09}; 
(8) \cite{vandyk12a};
(9) \cite{bose13};
(10) \cite{maund13}; (11) \cite{barbarino15}.
}
\end{table*}

Our code produces bolometric LCs, so we need to compute bolometric luminosities for our data set. We use the correlation between bolometric correction and colours inferred by \cite{bersten09}, which allow us to calculate bolometric luminosities using only two optical filters. In this work we calculated bolometric luminosities from \emph{BVI} photometry.

The observed photospheric velocity needs to be estimated through the measurement of certain spectroscopic lines. We use the Fe\,{\sc ii} ($\lambda$ 5169\,\AA) line since this line is formed in internal regions of SNe and has been proposed as a good estimator of the photospheric velocity \citep{dessart05}. 

\begin{figure}
\centering
        \includegraphics[width=0.51\textwidth]{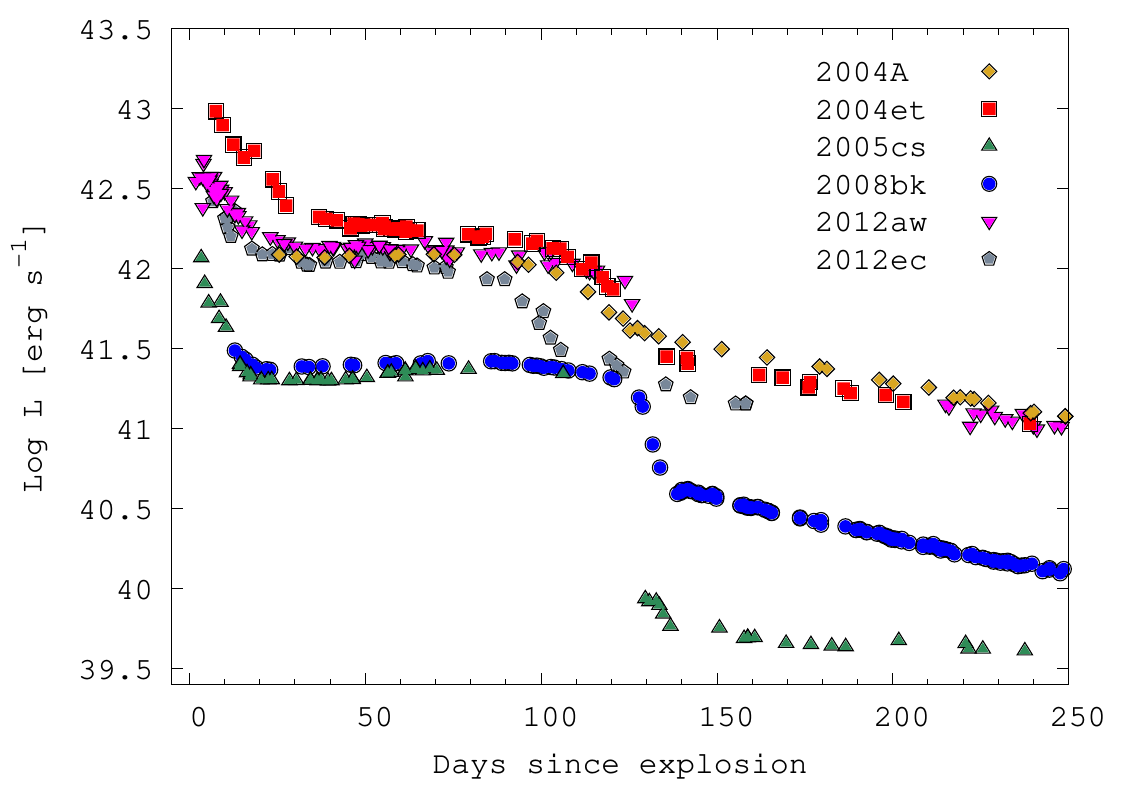}
        \begin{flushleft}
                \caption{Bolometric LCs of our SN sample.}
                \label{fig:sample_lc}
        \end{flushleft}
\end{figure}

\begin{figure}
\centering
        \includegraphics[width=0.5\textwidth]{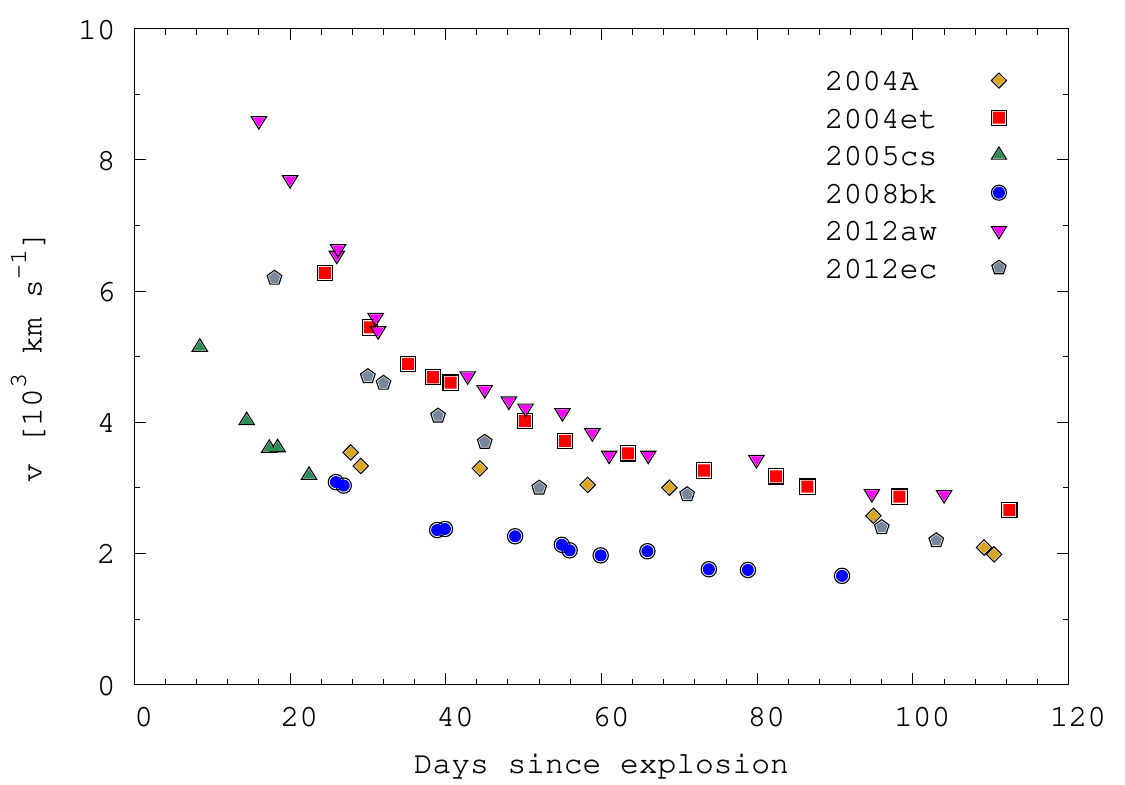}
        \begin{flushleft}
                \caption{Photospheric velocity evolution of our SN sample. We use the Fe\,{\sc ii}  $\lambda$ 5169\,\AA \,line as an indicator of the photospheric velocity.}
                \label{fig:sample_velo}
        \end{flushleft}
\end{figure}

In Figs. \ref{fig:sample_lc} and \ref{fig:sample_velo} we compare bolometric LCs and velocity evolution of our sample of SNe. We note that SNe 2004A, 2004et, and 2012aw are the most luminous, present the highest expansion velocities, and have synthesised the largest amount of nickel during explosion, since the bolometric luminosity of the radioactive tail is an almost direct indicator of the amount of nickel produced. On the other hand, SNe 2005cs and 2008bk are the faintest ones and also those that have the lowest velocities. Interestingly, these two SNe show similar LCs except during the radioactive phase, where SN 2008bk is substantially more luminous. Thus, we can see that even though the sample is small, there is a large variety of plateau luminosities, durations, tail luminosities, and expansion velocities.

\subsection{SN 2004A}
\label{subsec:04a}

SN 2004A was discovered by K. Itagaki (Teppo-cho, Yamagata, Japan) using a 0.28-m f/10 reflector on 2004 January 9.84 UT and later confirmed on 2004 January 10.75 UT \citep{nakano04}. SN 2004A was located at RA $=$ 16$^{h}$43$^{m}$01.90$^{s}$, Dec. $=$ +36$^{\circ}$50$'$12.5$''$ (equinox 2000.0), around 22$''$ west and 17$''$ north of the centre of the nearby spiral galaxy NGC 6207. Optical spectra obtained on 2004 January 11.8 UT and 11.9 UT showed a blue continuum and the hydrogen Balmer lines with P-Cygny profiles \citep{kawakita04}. The emission components were somewhat weak suggesting that it was indeed a young type II SN.

Nothing was visible at this position in the observations of  Itagaki on 2003 December 27. Although the discovery was close to the date of last observation, this is not enough to have a precise estimation of the explosion epoch ($t_{\rm exp}$) since there is an uncertainty of $\sim$10 days. \cite{hendry06} estimated the explosion epoch from the comparison of the optical LC with that of SN 1999em giving a value of JD 245\,3011 $\pm$ 3, that is 4 days before the discovery. Assuming this value for the explosion epoch we were not able to find a set of parameters that reasonably reproduces the observations. The method used to estimate $t_{\rm exp}$ is not precise. It is based on LC comparison with other SN that do not necessarily have the same LC evolution. Therefore, we decided to modify $t_{\rm exp}$ based on our modelling. However, the value adopted (see Table \ref{table:info}) is marginally outside of the value proposed by \citet{hendry06} considering the error bars.

There are archival pre-explosion images of the site taken with the Wide Field and Planetary Camera 2 (WFPC2) on board the Hubble Space Telescope (HST) on 2000 August 03 and 2001 July 02. Observations were conducted in the three filters F300W, F606W, and F814W. \cite{hendry06} presented an analysis of these pre-explosion observations in conjunction with post-explosion HST Advanced Camera for Surveys (ACS) observations of the SN acquired on 2004 September 23. A progenitor candidate was identified in pre-explosion WFPC2 F814W images, but no object was visible in either the F606W or the F300W frames at the position of SN 2004A. These latter authors propose that the progenitor is a RSG with a possible main-sequence mass of 9$^{+3}_{-2}$~$M_{\odot}$. \cite{maund14a} presented late-time HST ACS observations of the site confirming the progenitor identification through its disappearance.

\subsection{SN 2004et}
\label{subsec:04et}

SN 2004et was discovered in the nearby galaxy NGC 6946 by S. Moretti on 2004 September 27 using a 0.4m telescope \citep{zwitter04}. This object is located at RA $=$ 20$^{h}$35$^{m}$25$^{s}$.33, Dec. $=$ +60$^{\circ}$07$'$17$"$.7 (equinox 2000.0). A high-resolution spectrum taken on 2004 September 28 with the Mt. Ekar 1.82m telescope showed a relatively featureless spectrum with a very broad, low-contrast H$\alpha$ emission, classifying it as a type II SN.

This is the case of a SN for which there is a very good determination of the explosion epoch because there are no detections in the site of the SN on images taken a few hours before explosion. K. Itagaki (Teppo-cho, Yamagata, Japan) found nothing at the location of SN 2004et on 2004 September 19.655 UT. On 2004 September 22.017 UT, the robotic telescope TAROT reported \textit{R}-band magnitudes but nothing was visible to a limiting magnitude of 19.4 $\pm$ 1.2. On 2004 September 22.983, in the same site, there was a 15.17 $\pm$ 0.16 mag detection \citep{yamaoka04}. The explosion epoch is therefore well constrained and is taken as 2004 September 22.0, that is JD 245~3270.5 \citep{li05}.

The SN 2004et was also detected in X-rays and radio waves suggesting the presence of substancial circumstellar material (CSM) around the SN \citep{stockdale04,misra07}. The interaction between the material ejected by the SN and the CSM created a shocked region producing X-ray and radio synchrotron emission. In fact, three years after the explosion, the emission line profiles of spectra taken in the mid-infrared still indicated the existence of interaction \citep{kotak09}.

The search for the progenitor in archival pre-explosion Canada-France-Hawaii Telescope (CFHT) images led \cite{li05} to identify a yellow supergiant star as a candidate progenitor. These images included \emph{BVR} images from 2002 and \emph{u$'$g$'$r$'$} observations from 2003. However, three years after explosion, when this SN had faded sufficiently to allow verification of whether or not the candidate progenitor had disappeared, \cite{crockett11} showed that the source indicated as the progenitor was still visible in observations from the William Herschel Telescope. High-resolution HST WFPC2 and Gemini North images revealed that this source is resolved into at least three distinct sources. Also, \cite{crockett11} reported the discovery of the progenitor as an excess of pre-explosion flux in the \emph{R}- and \emph{I}-band. By combining the \emph{R}- and \emph{I}-band photometry these latter authors estimated that the progenitor is a RSG with a tentative ZAMS mass of 8$^{+5}_{-1}$~$M_{\odot}$.
However, they stress that future high-resolution observations will be required to confirm or adjust this result, since the progenitor photometry would require revision if the SN were found to have faded still further.

\subsection{SN 2005cs}
\label{subsec:05cs}

This SN was discovered on 2005 June 28.905 UT in the galaxy M51 \citep{kloehr05} and spectroscopically classified as a young type II SN due to its blue continuum and P-Cygni profiles of the Balmer and He lines \citep{modjaz05}. Kloehr gave the position for SN 2005cs as RA $=$ 13$^{h}$29$^{m}$53$^{s}$.37, Dec. $=$ +47$^{\circ}$10$'$28$''$.2 (equinox 2000.0), which is 15$''$ west and 78$''$ south of the centre of the host galaxy.

The earliest detection was made by M. Fiedler on 2005 June 27.91 UT. Nothing was visible on 2005 June 20.598 UT (Kloehr et al. 2005) at the SN position. In fact, images obtained on 2005 June 26 by amateur observers have shown no detection either. In particular, the SN site was monitored on 2005 June 26.89 using a 0.4m telescope and nothing was detected below the magnitude limits. Therefore, these limitations in the detection restrict the uncertainty in the explosion epoch to only one day.

Pre-explosion images of the site were available in the HST archive dating from 2005 January, taken with the ACS instrument in four bands: F435W, F555W, F814W, and F658N. The ACS data provide the deepest and the highest-resolution optical images currently available of the galaxy prior to the SN. The galaxy M51 was also observed by the Near Infrared Camera and Multi-Object Spectrometer (NICMOS) on board HST in 1998 in five bands, and with the Gemini North Telescope Near InfraRed Imager (NIRI) in the \emph{JHK} bands in 2005 April. In addition to these, the region was observed with the WFPC2 in 1999 using three filters: F336W, F555W, and F675W \citep{maund05,li06}. In spite of the extensive data available, the progenitor candidate was only detected in the F814W band. Based on this and on upper limits in the other bands, \cite{maund05} and \cite{li06} determined that the progenitor is a RSG star of spectral type K3 or later with an initial mass between 7 and 13~$M_{\odot}$.

Through late-time HST ACS WFC observations of the site, \cite{maund14a} confirmed the progenitor identification in virtue of its disappearance.

SN 2005cs belongs to a group of underluminous SNe II. It has been suggested that their progenitors are close in mass to the lower limit for stars which can undergo core-collapse \citep{chugai00}.

\subsection{SN 2008bk}
\label{subsec:08bk}

This SN was discovered on images taken on 2008 March 25.13 UT. It is located at RA = 23$^{h}$57$^{m}$47$^{s}$.5, Dec. = -32$^{\circ}$33$'$24$''$ (equinox 2000.0), which is 26" east and 138" north of the nucleus of the galaxy NGC 7793 \citep{monard08}. Nothing was visible at this position on images taken by Monard on 2008 January 2.742 UT. As there are no images closer to the explosion epoch, this is not well determined. In fact, we found two different estimations of the explosion epoch in the literature. \cite{morrell08} classified this SN as a type II with an age of 36 days after explosion on 2008 April 12.4 (JD 245\,4532), based on a comparison with the well-studied SN 1999em. On the other hand, G. Pignata, (private communication) determined the explosion epoch by comparing the optical light curve with that of SN2005cs and obtained JD 245\,4548 $\pm$ 2. Given that the methods to determine the explosion epoch are not consistent, we use a different value of JD 245\,4543 based on our modelling, which is intermediate between the other two estimates.

\cite{li08} were the first to identify the progenitor star as a RSG close to the position of SN 2008bk in deep archival ground-based \emph{BVI} pre-explosion images obtained in 2001 with one of the 8.2\,m telescopes of the European Southern Observatory (ESO). A full identification of the progenitor was given by \cite{mattila08} using high-quality optical and near-infrared (NIR) pre-explosion images from the Very Large Telescope (VLT), concluding that the progenitor of SN 2008bk is a RSG with an initial mass of 8.5 $\pm$ 1.0~$M_{\odot}$. Moreover, \cite{vandyk12a} measured accurate photometry for the RSG progenitor from superior-quality \emph{g$'$r$'$i$'$} images obtained in 2007 with the Gemini-South 8\,m telescope, as well as from the NIR VLT archival images, and concluded that the progenitor is a RSG with initial mass in the range of 8$-$8.5~$M_{\odot}$. On the other hand, \cite{maund14b} found that the progenitor was a highly reddened RSG and estimated an initial mass of 12.9$^{+1.6}_{-1.8}$~$M_{\odot}$. The progenitor of SN 2008bk, detected in six optical bands, is the second-best-characterised progenitor to date, after that of SN 1987A. In 2011, deep images of the site of explosion were taken confirming the disappearance of the progenitor \citep{mattila10, vandyk13}.

\subsection{SN 2012aw}
\label{subsec:12aw}

The SN 2012aw was discovered on 2012 March 16.9 UT in the galaxy M95 \citep{fagotti12}. On 2012 March 15.3, the site of the explosion had been observed without detections, so the explosion epoch is well constrained. \citet{fraser12} set the explosion epoch to 2012 March 16.0 $\pm$ 0.8 UT, corresponding to JD 245\,6002.5. Several spectra taken in the following days were used to classify it as a type II SN \citep{manuri12,siviero12}.

Pre-explosion images were available of the SN site obtained with the HST  WFPC2 camera between 1994 and 2009. Observations were conducted in five filters: F336W, F439W, F658N, F555W, and F814W. Images have also been found in the ESO archive taken with NTT+SOFI in \emph{K$_{s}$}-band and VLT+ISAAC in the \emph{J$_{s}$}-band between 2000 and 2006. From these observations, a source could be detected in four bands at the SN location. The progenitor was determined as a RSG star with initial mass range of 14$-$26~$M_{\odot}$ \citep{fraser12,vandyk12b}. Three years after the explosion, \cite{fraser16} confirmed the progenitor identification through its disappearance.

\subsection{SN 2012ec}
\label{subsec:12ec}

SN 2012ec was discovered in the galaxy NGC 1084 on 2012 August 11.039 UT \citep{monard12}. A spectrum of the SN acquired the following day showed it to be a young type II a few days post-explosion \citep{childress12}. As in the case of SN 2008bk, there are no previous images close to the explosion, and so the estimation of the explosion epoch is not precise. It was estimated as JD 245\,6143.5 based on comparisons with spectra of SN 2006bp \citep{barbarino15}. As we have mentioned before, we find this method unreliable. In any case, comparing our models with observations, we could not find a set of parameters that describe LC and velocity evolution well enough. Therefore, we decided to use a value more consistent with our modelling (see Table \ref{table:info}).

There are pre-explosion images from 2001 taken with the WFPC2 on board HST in three bands F450W, F606W, and F814W. In the HST archive, images taken in 2011 were found but only in F814W-band. \cite{maund13} were able to detect a progenitor candidate in the WFPC2 F814W image, inferring an initial mass range of 14$-$22~$M_{\odot}$. Late-time observations confirmed the progenitor identification through its disappearance (S. Van Dyk, private communication).

\section{Hydrodynamic models} 
\label{sec:models}

Theoretical LCs are calculated using a 1D Lagrangian hydrodynamical code that simulates the explosion of the SN and produces bolometric light curves and photospheric velocities of SNe II \citep{bersten11}. The code solves numerically the hydrodynamical equations assuming spherical symmetry for a self-gravitating configuration. Radiation transport is treated in the diffusion approximation for optical photons and grey transport for gamma photons produced by the radioactive decay of \Ni. The explosion is simulated by injecting a certain amount of energy near the centre of the progenitor for a short time as compared with the hydrodynamic timescale. This energy induces the formation of a powerful shock-wave that propagates through the progenitor transforming thermal and kinetic energy of the matter into energy that can be radiated from the stellar surface.

Several approximations are made in the equations of radiation hydrodynamics. The code assumes that the fluid motion can be described as a 1D, radially symmetric flow. This might not be entirely correct since it is assumed that the explosion mechanism of core-collapse SNe (CCSNe) may be a highly asymmetric process. However, \citet{leonard05} showed that due to the very extended hydrogen envelopes that characterise SNe II-P, the asymmetries expected from the explosion mechanism itself appear to be smoothed, meaning that spherical symmetry is a good approximation for the bulk of the ejecta.

The code adopts local thermodynamical equilibrium (LTE) to describe the radiative transfer. This approximation assumes that radiation and matter are strongly coupled, which is not valid at shock breakout and during and after the transition phase from optically thick to optically thin ejecta, when this is completely recombined.

The code uses opacity tables calculated assuming LTE and a medium at rest. These calculations underestimate the true line opacity when considering rapidly expanding envelopes where large velocity gradients are present \citep{karp77}. Another effect that is not included in the calculation of the opacity is the non-thermal excitation or ionisation of electrons that are created by Compton scattering of $\gamma$-rays emitted by radioactive decay of \Ni\, and $^{56}$Co. The LTE ionisation used in the calculation of the opacity considerably underestimates the true ionisation. To partially solve the underestimation in the mean opacity, the code adopts an alternative approximation that has been tested in the literature, which consists in using a minimum value of the opacity sometimes referred to as the opacity floor \citep[see][for details]{bersten11}. This approximation may introduce the largest uncertainties in the derived parameters. However, a quantitative evaluation of its effects is beyond the scope of this work.

A pre-supernova model in hydrostatic equilibrium that simulates the conditions of the star before exploding is necessary to initialise the explosion. Two different types of initial (or pre-SN) models are typically employed in the literature: those coming from stellar-evolution calculations (or “evolutionary” models), and those from non-evolutionary calculations (or “parametric” models) where the initial density and chemical composition are parameterised in a convenient way. In this work we use double polytropic models in hydrostatic equilibrium as non-evolutionary pre-SN models to make a complete description of the physical parameters of the sample. The motivation behind our choice of double polytropic models is provided below. In Appendix \ref{ap:stellar_evolution}, the initial density profile used for each SN in our sample is shown, together with a comparison of our results for SN 2008bk using evolutionary models.

To determine physical parameters for a given SN we compare different models with observations. The free parameters of the model are: mass and radius of the progenitor ($M_{\rm hydro}$ and $R$), the energy that is transferred to the envelope after core-collapse (denoted as “explosion energy”; $E$) and the amount of radioactive material synthesised in the explosion ($M_{\rm Ni}$) and its degree of mixing into the outer layers in the ejecta. For parametric models, the progenitor mass and radius can be treated as independent parameters, as opposed to the evolutionary models. This is the main motivation to use double polytropic models as pre-SN structures. 

It is important to mention that there is a degree of degeneracy between the progenitor mass, radius, and explosion energy. This can be partially reduced by modelling LCs together with the photospheric velocity evolution, and is ideal if also modelling the spectra. To reduce the degeneracy  even more we decided to model those observables treating the progenitor radius as a fixed parameter with a value derived by the pre-explosion data. We would like to point out that the progenitor radius, and not the mass
which depends on an evolutionary model, is the most direct progenitor parameter that can be derived using the pre-explosion data. Progenitor detections in pre-explosion images give a measure of the spectral energy distribution (SED). This allows us to determine effective temperature ranges, and a value for the progenitor radius (assuming a black body) can be estimated  by calculating
the bolometric luminosity \citep[][among others]{vandyk12a,maund13}. 

Our goal is then to test if it is possible to find a good representation of the LC and the photospheric velocities of each SN of our sample using the values of the radius determined by direct detections. 
In most cases, more than one value for the radius was determined from the analysis of archival images, because there are different values given by different authors, or due to the range of derived values of $L$ and $T_{eff}$. Table \ref{table:radios_preexp} shows the range of the progenitor parameters (main sequence mass and pre-SN radius) for each SN in our sample derived by different authors using the pre-explosion data. In cases where there is more than one estimation of the pre-SN radius, we have decided to take the complete range of values predicted in the different works instead of using some specific value because we do not have sufficient information to prioritise one value over another.

An important point to clarify is the meaning of the different masses that we consider here. The mass used in the polytropic models refers to the mass of the star just before the explosion (which we call hydrodynamic mass; $M_{\rm hydro}$). This value is not necessarily the same as the value of the mass of the star in the zero age main sequence ($M_{\rm \,ZAMS}$). Moreover, $M_{\rm hydro}$ is usually smaller than $M_{\rm \,ZAMS}$ since the star loses mass during its evolution. On the other hand, the masses that are derived by direct detections in pre-explosion images do refer to the masses of the stars in the ZAMS. Therefore,  we must keep this in mind when comparing both methods.

Another parameter to consider is the ejecta mass ($M_{\rm ej}$), which is equal to the mass of the pre-SN object minus the mass of the compact remnant forming during core collapse. In all of our calculations, we assume that the mass of the compact remnant is 1.4~$M_{\odot}$. Therefore this part is removed from the explosion.

\begin{table}
        \caption{Range of values for the mass and radius of the progenitors in our SN sample derived through direct detections in pre-explosion images by differents authors. Only the radius information is used in our hydrodynamical modelling.}
        \label{table:radios_preexp}    
        \centering                
        \begin{tabular}{c c c c}      
                \hline\hline\noalign{\smallskip}    
                SN & M$_{\,\rm ZAMS}$  & Radius  & References \\
                   & [M$_{\odot}$]   & [R$_{\odot}$] & \\
                \hline\noalign{\smallskip}    
                2004A  & $7-14.3$   & $330-950$      & 1, 2, 3 \\
                2004et & $7-13$     & $350-980$      & 3, 4 \\
                2005cs & $6.6-13$   & $205-630$      & 3, 5, 6 \\
                2008bk & $7.5-14.5$ & $455-650$      & 3, 7, 8, 9, 10 \\
                2012aw & $11-26$    & $545-1140$     & 3, 11, 12, 13, 14 \\
                2012ec & $14-22$    & $1030 \pm 180$ & 3, 15 \\
        \hline
        \end{tabular}
        \tablebib{(1) \citet{hendry06}; (2) \citet{maund14a}; \linebreak (3) \citet{davies18};
        (4) \citet{crockett11};
        (5) \citet{maund05}; (6) \citet{li06};
        (7) \citet{mattila08}; (8) \citet{vandyk12a}; (9) \citet{maund14b}; (10) \citet{maund17};
        (11) \citet{vandyk12b}; (12) \citet{fraser12}; (13) \citet{kochanek12}; \linebreak (14) \citet{fraser16};
        (15) \citet{maund13}.
        }
\end{table}

\section{Results}
\label{sec:results}


\begin{figure*}
\centering
                \includegraphics[width=0.467\textwidth]{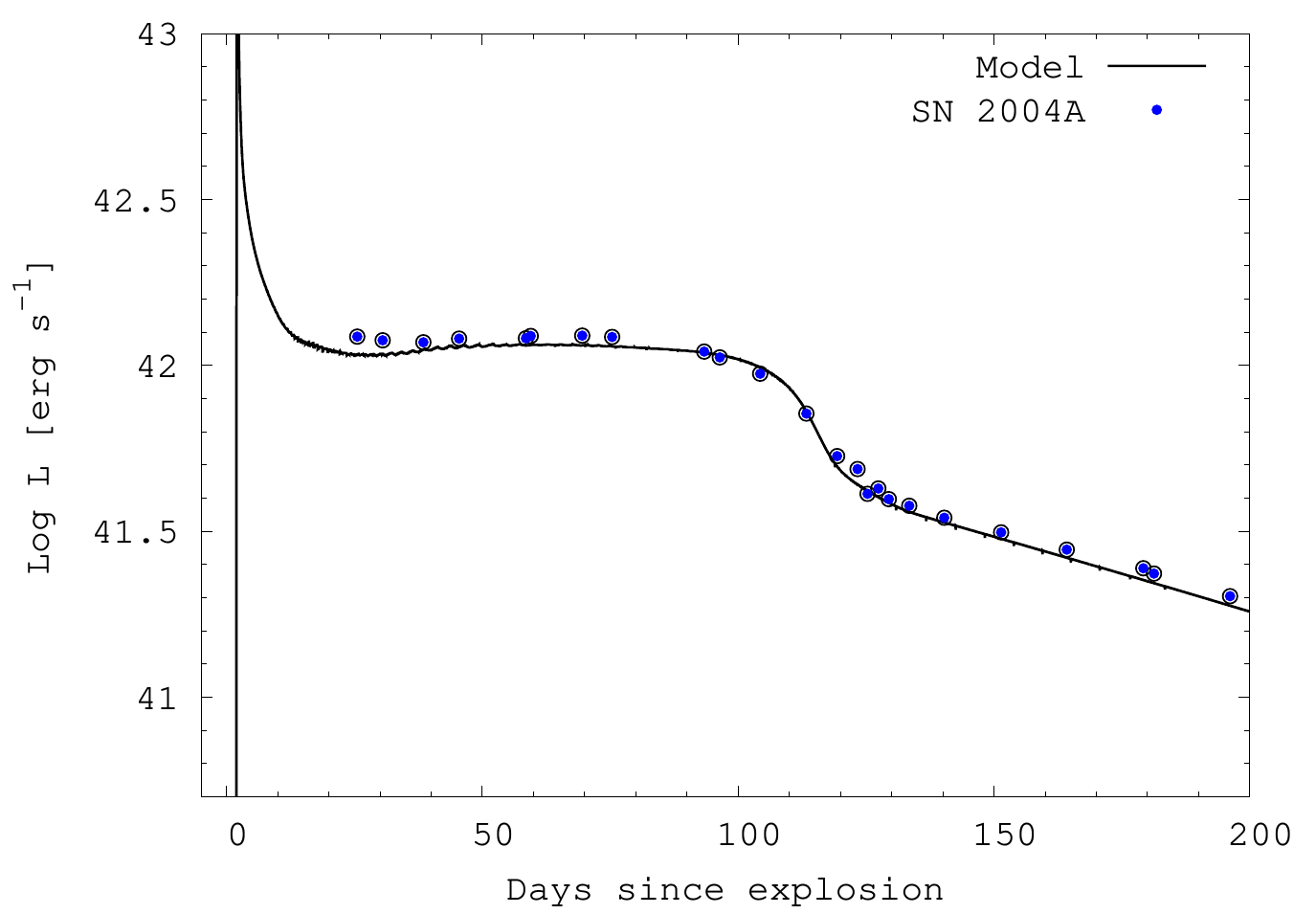}
                \includegraphics[width=0.467\textwidth]{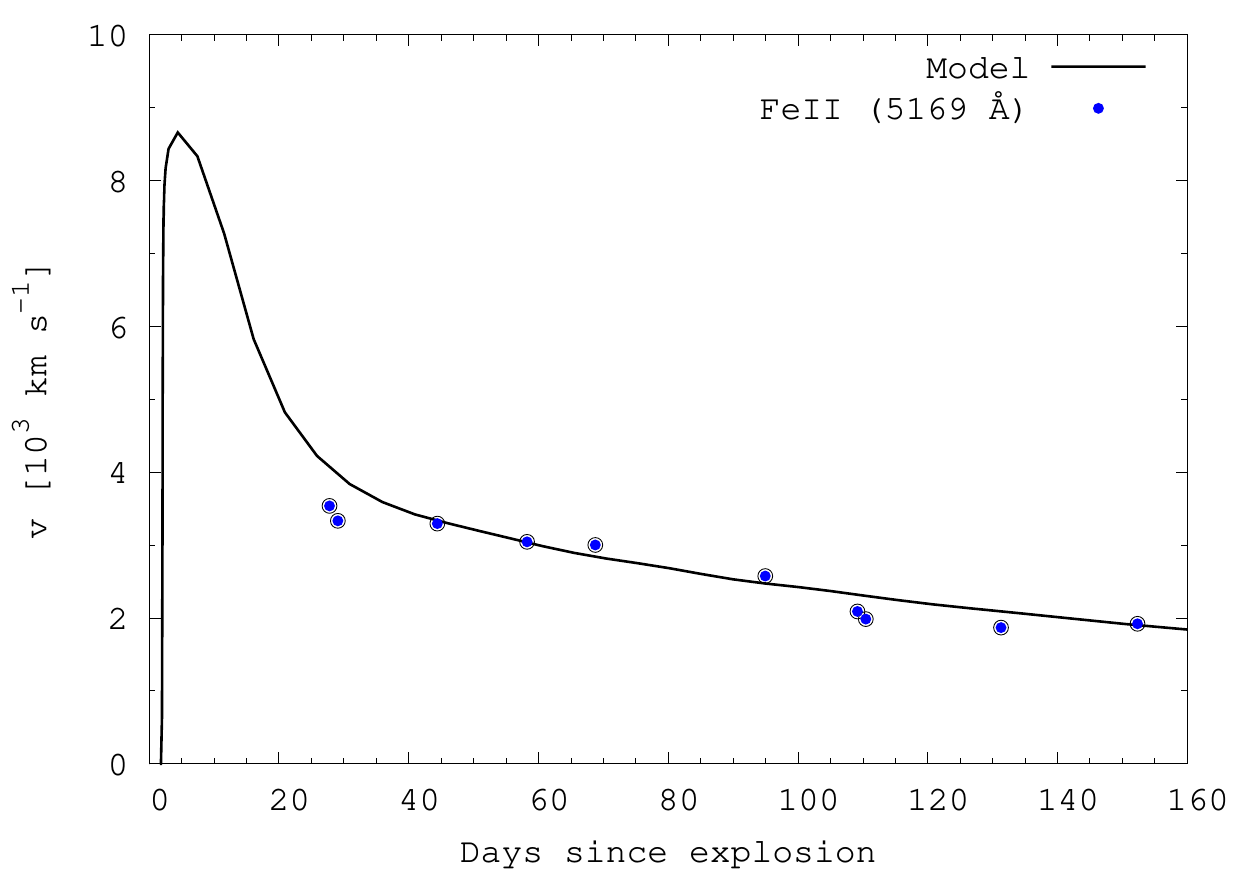}
                \includegraphics[width=0.467\textwidth]{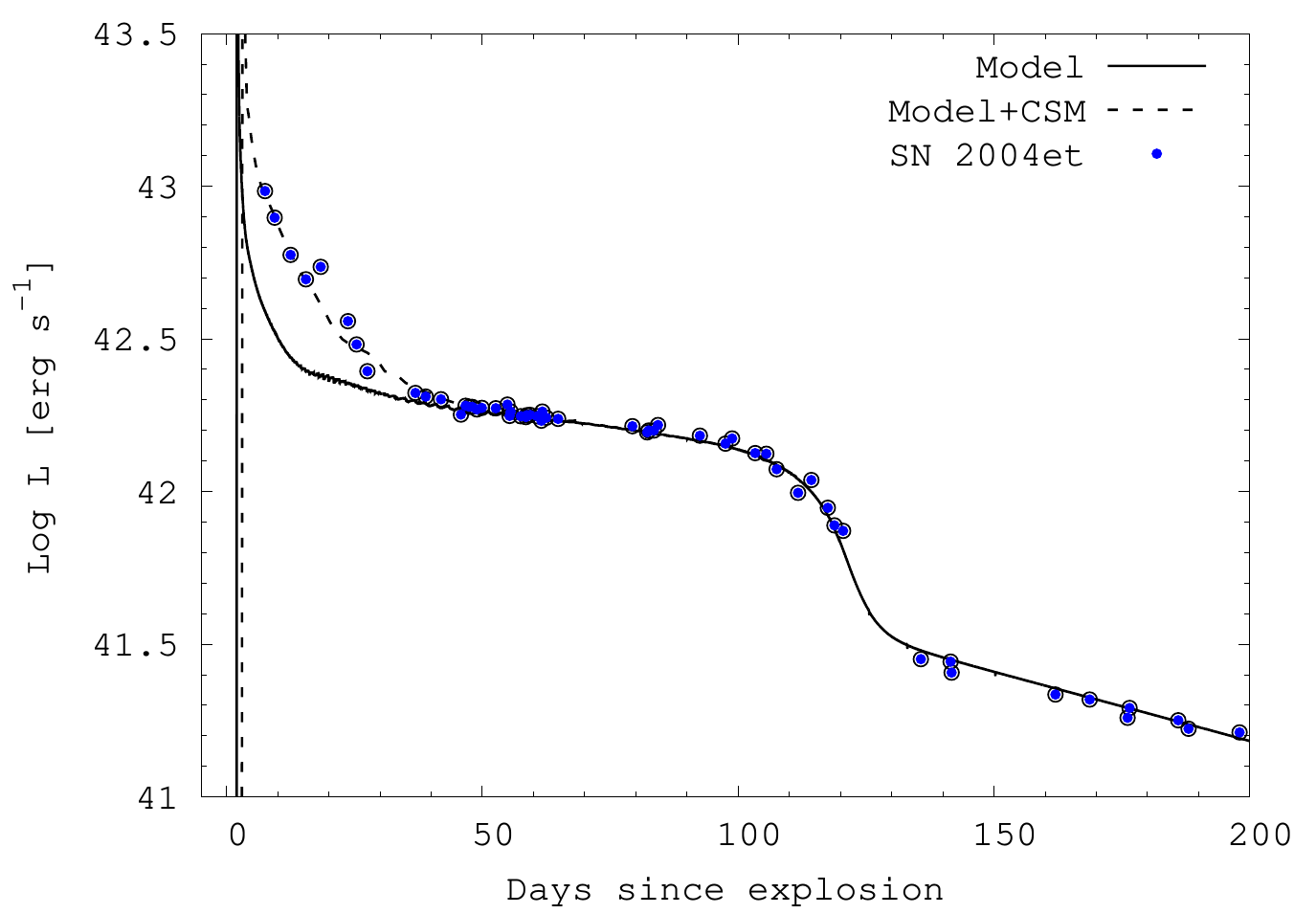}
                \includegraphics[width=0.467\textwidth]{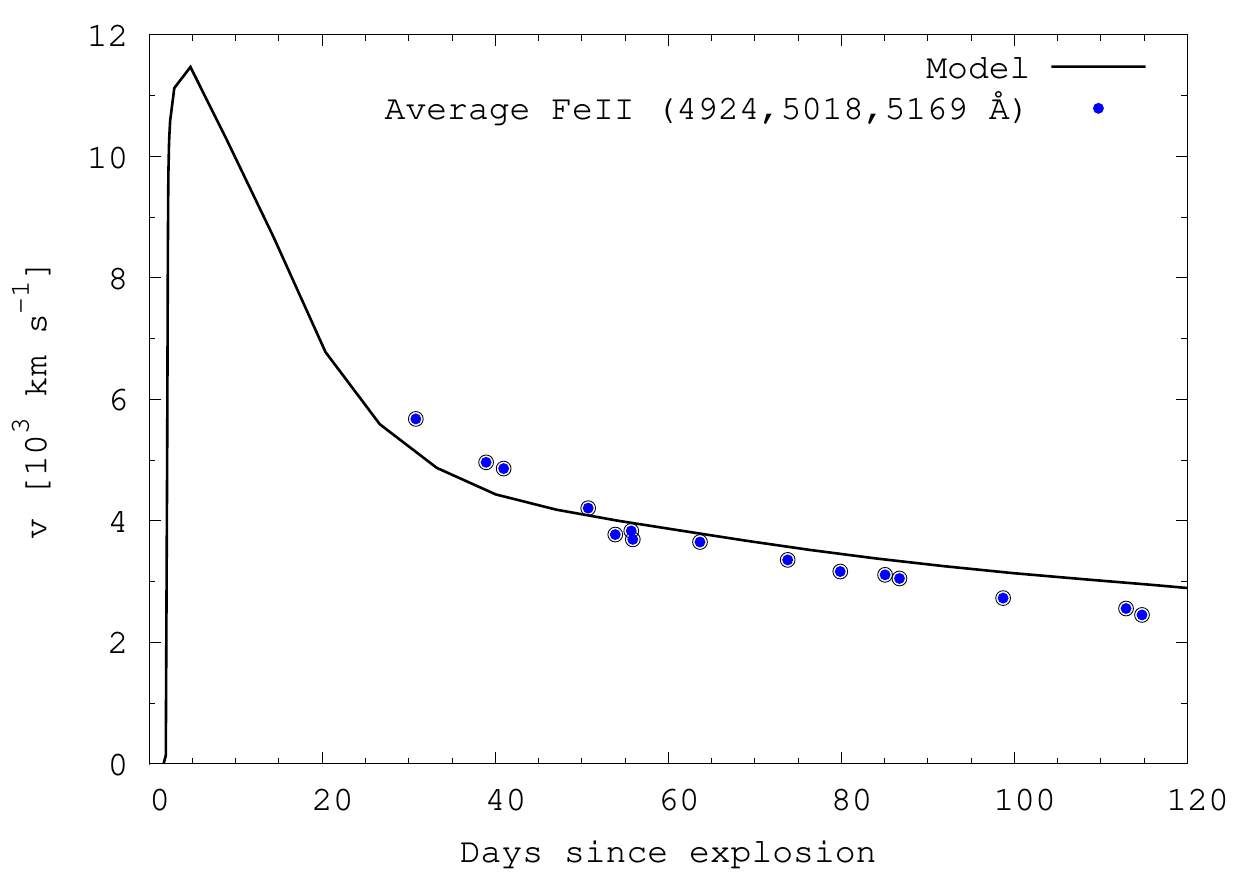}
                \includegraphics[width=0.467\textwidth]{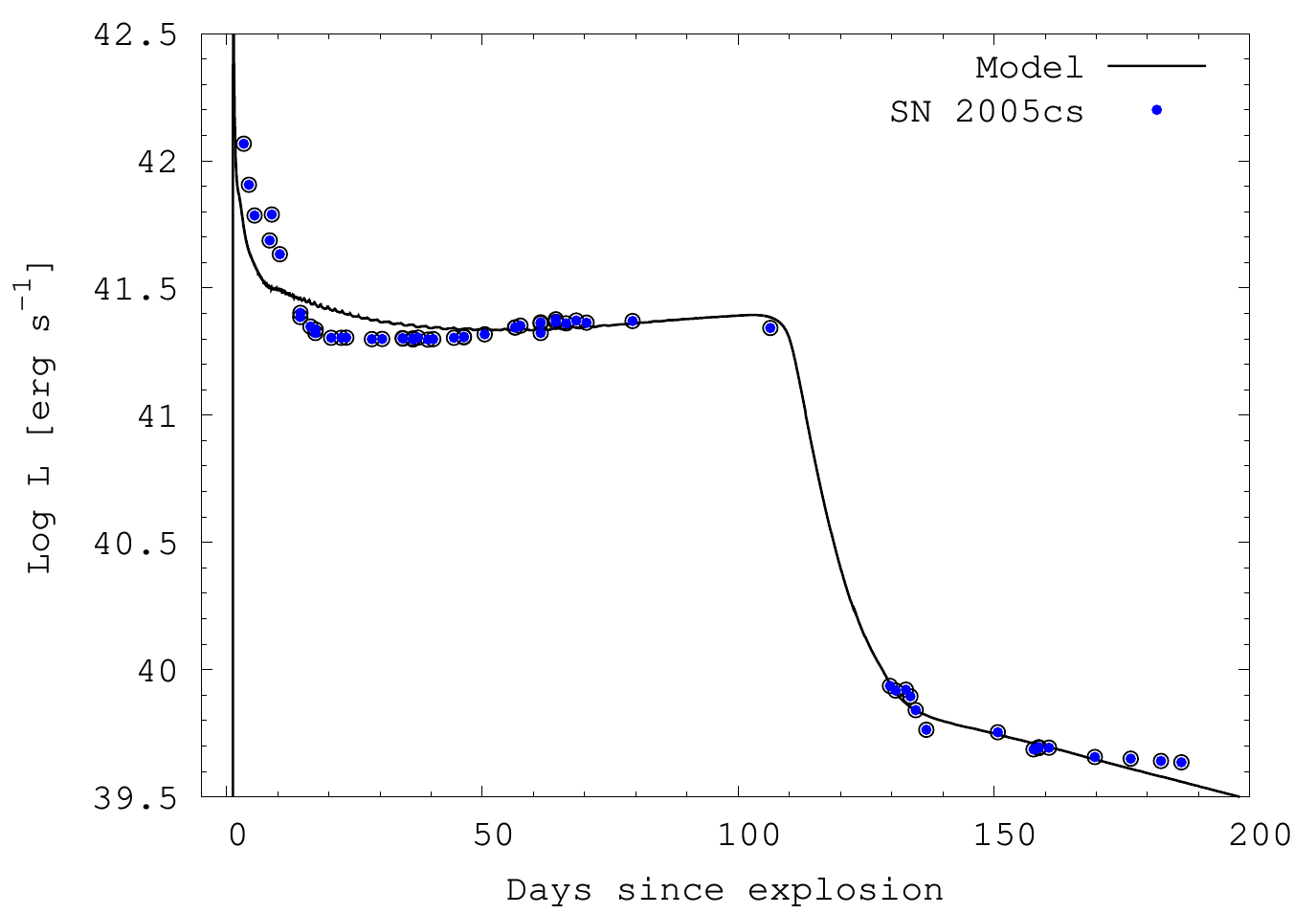}
                \includegraphics[width=0.467\textwidth]{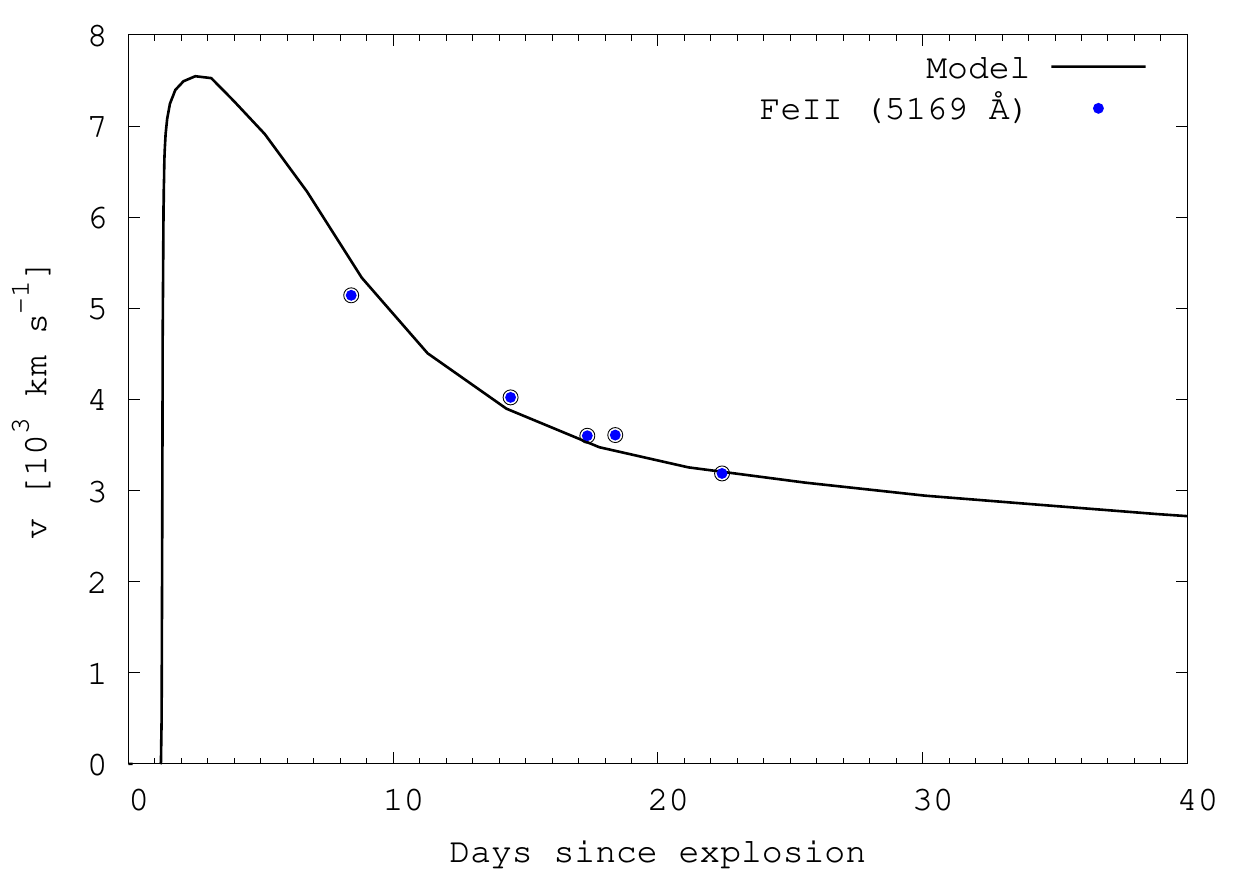}
                \includegraphics[width=0.467\textwidth]{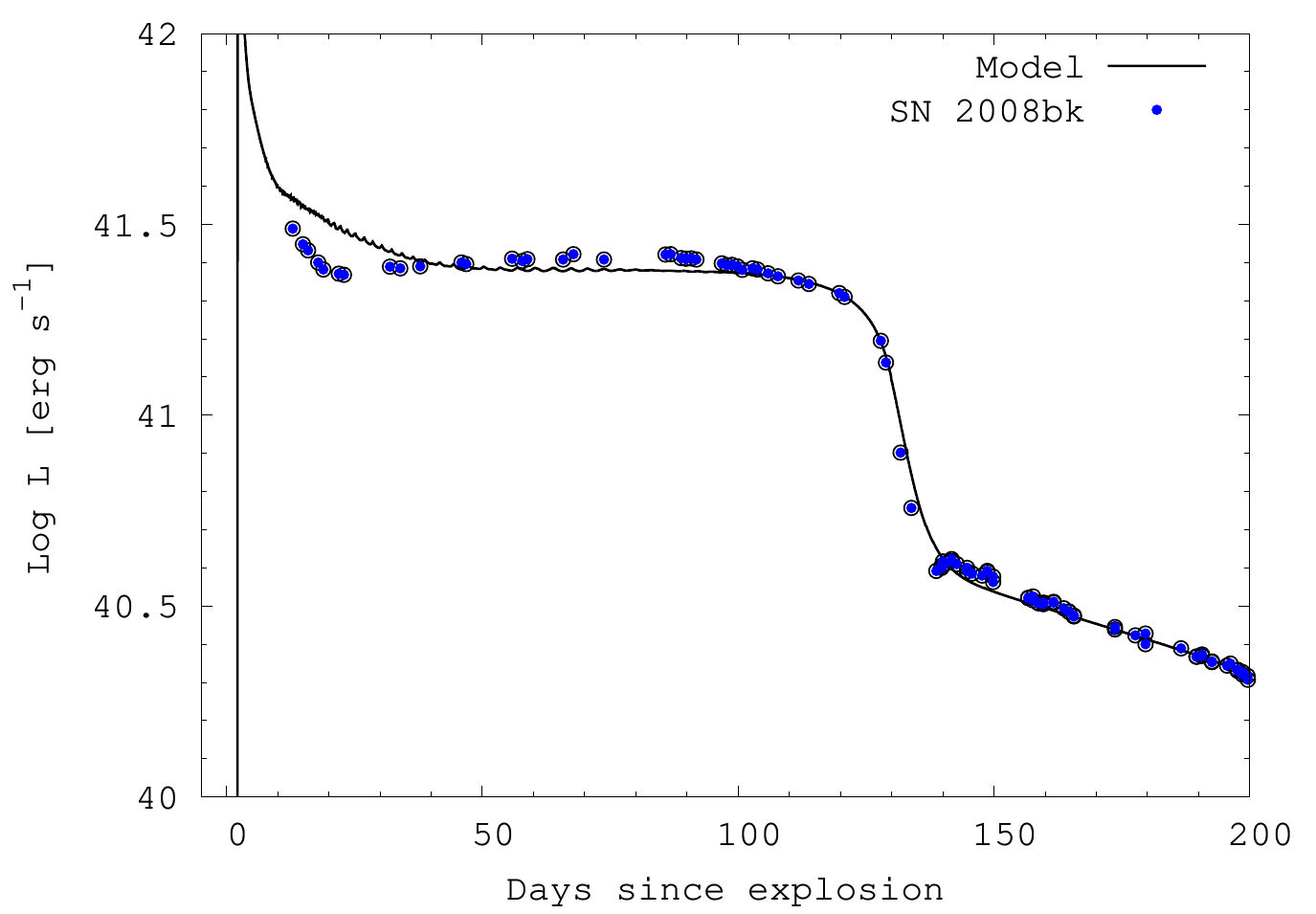}
                \includegraphics[width=0.467\textwidth]{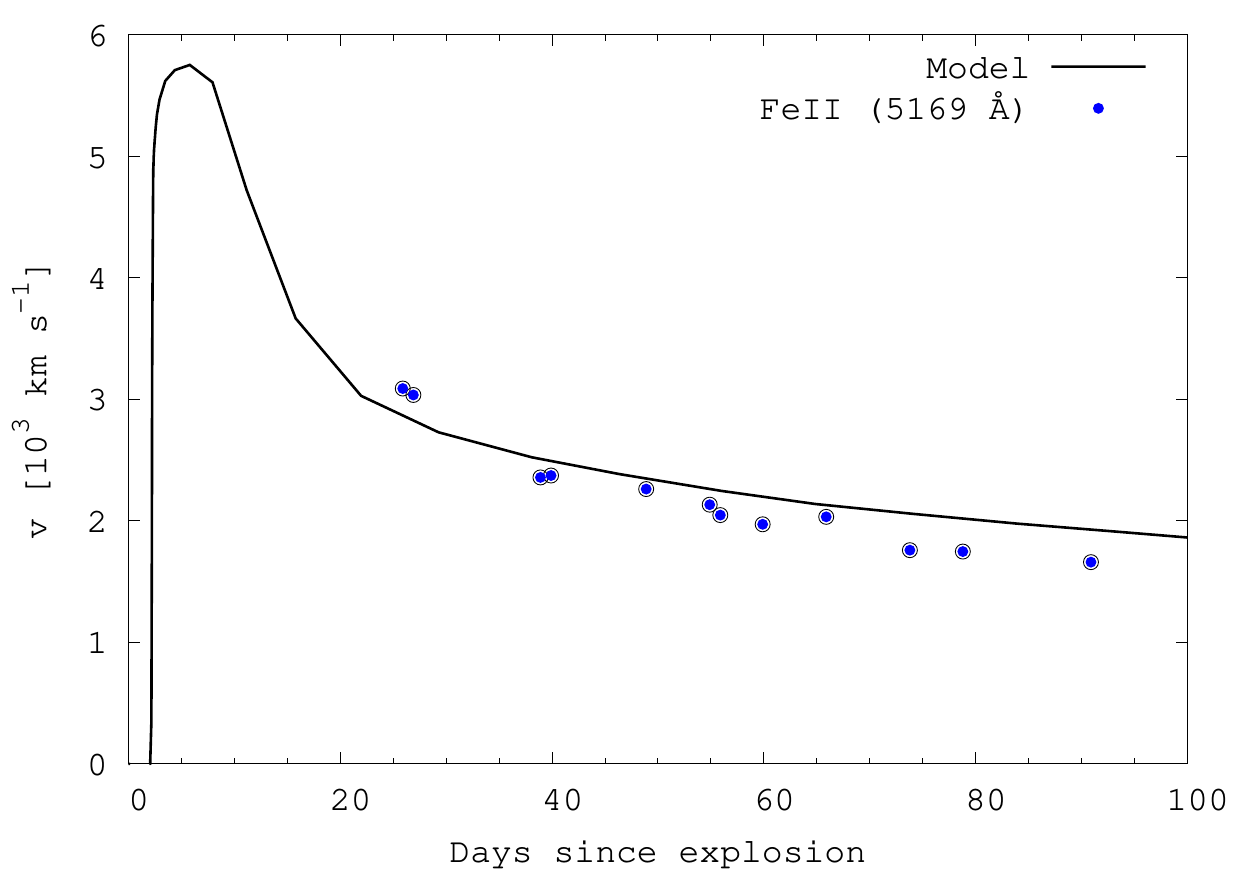}
        \begin{flushleft}
                \caption{Comparison between models and observations for our SN sample. (Left) Bolometric light curves. (Right) Evolution of the photospheric velocity. From top to bottom: SN 2004A, SN 2004et, SN 2005cs, and SN 2008bk.}
                \label{fig:models1}
        \end{flushleft}
\end{figure*}

\begin{figure*}
\centering
                \includegraphics[width=0.467\textwidth]{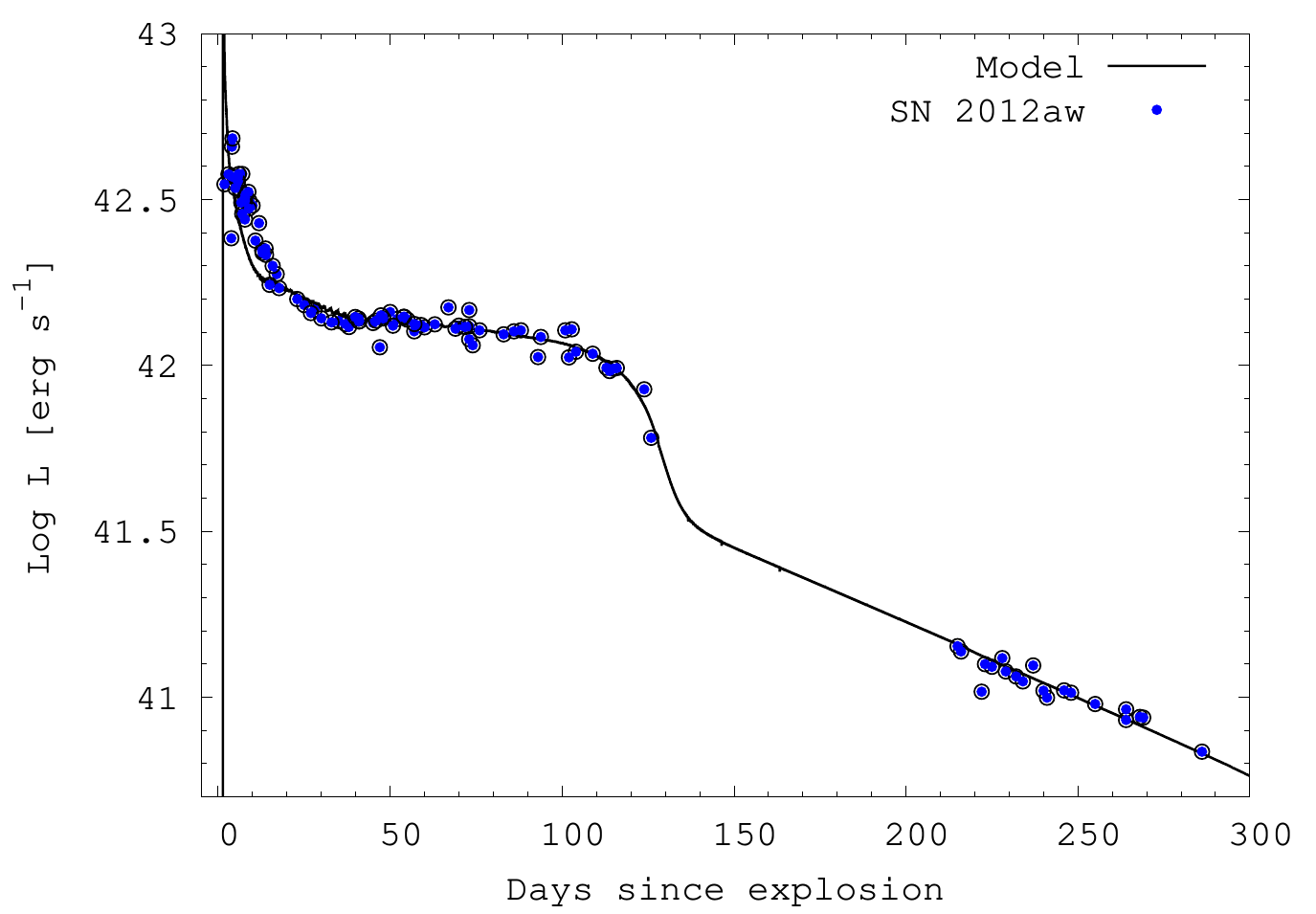}
                \includegraphics[width=0.467\textwidth]{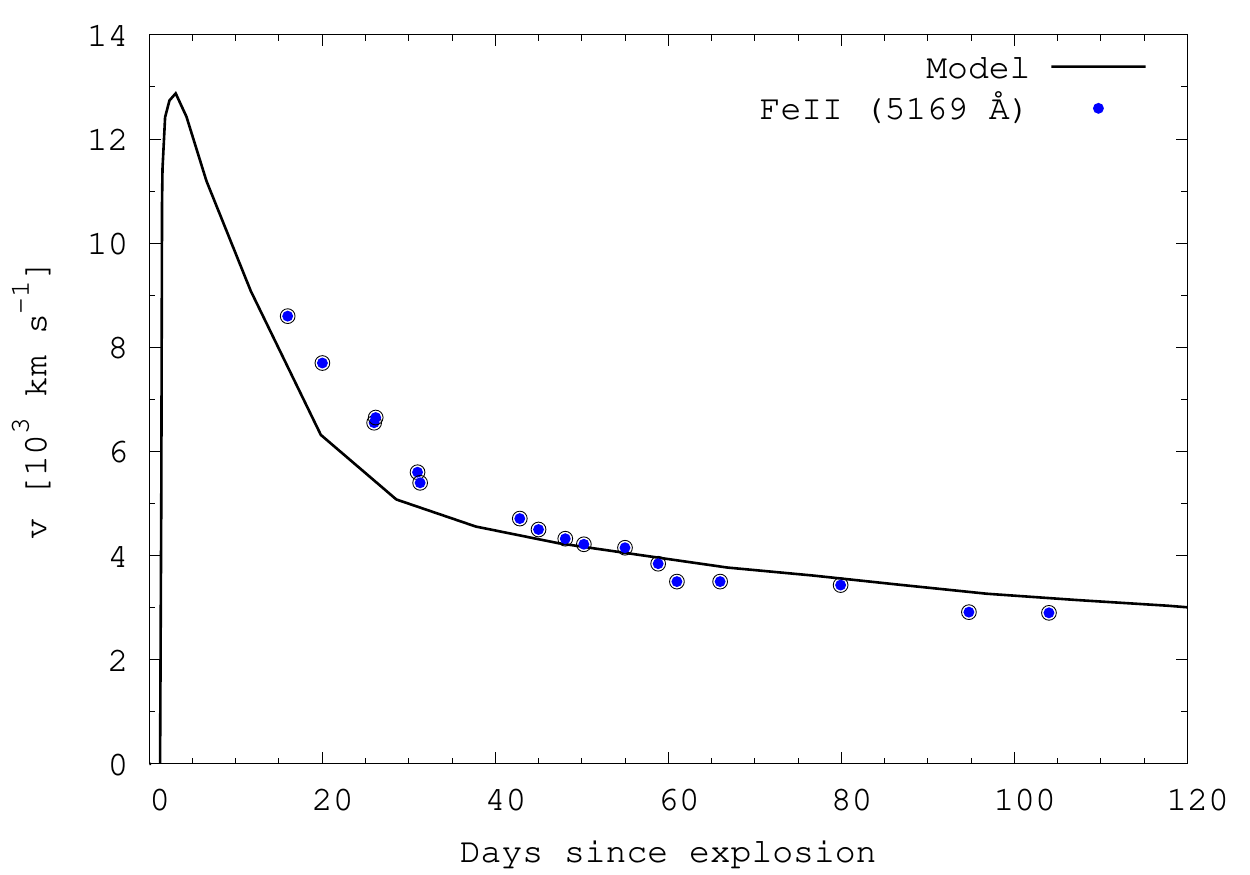}
                \includegraphics[width=0.467\textwidth]{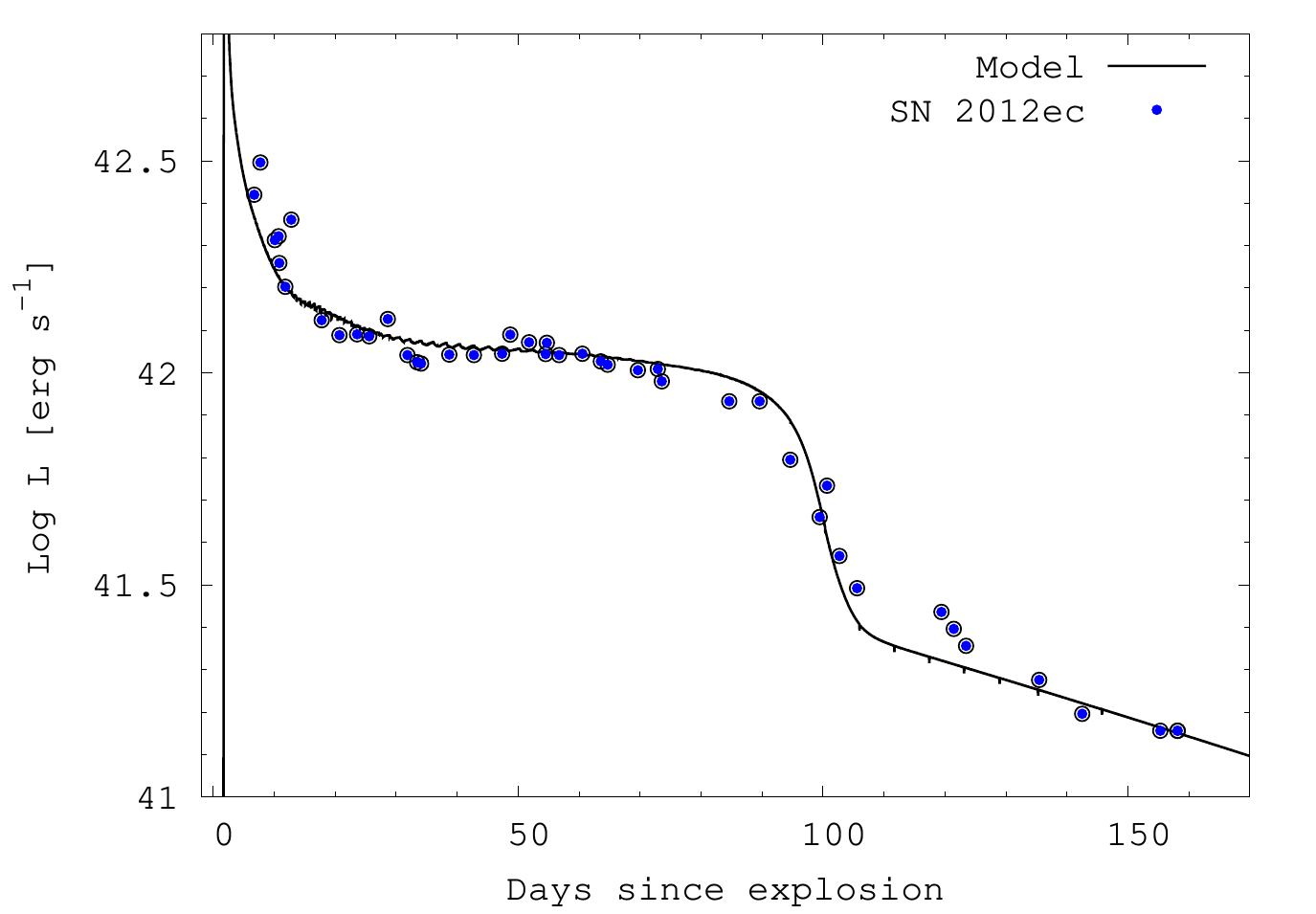}
                \includegraphics[width=0.467\textwidth]{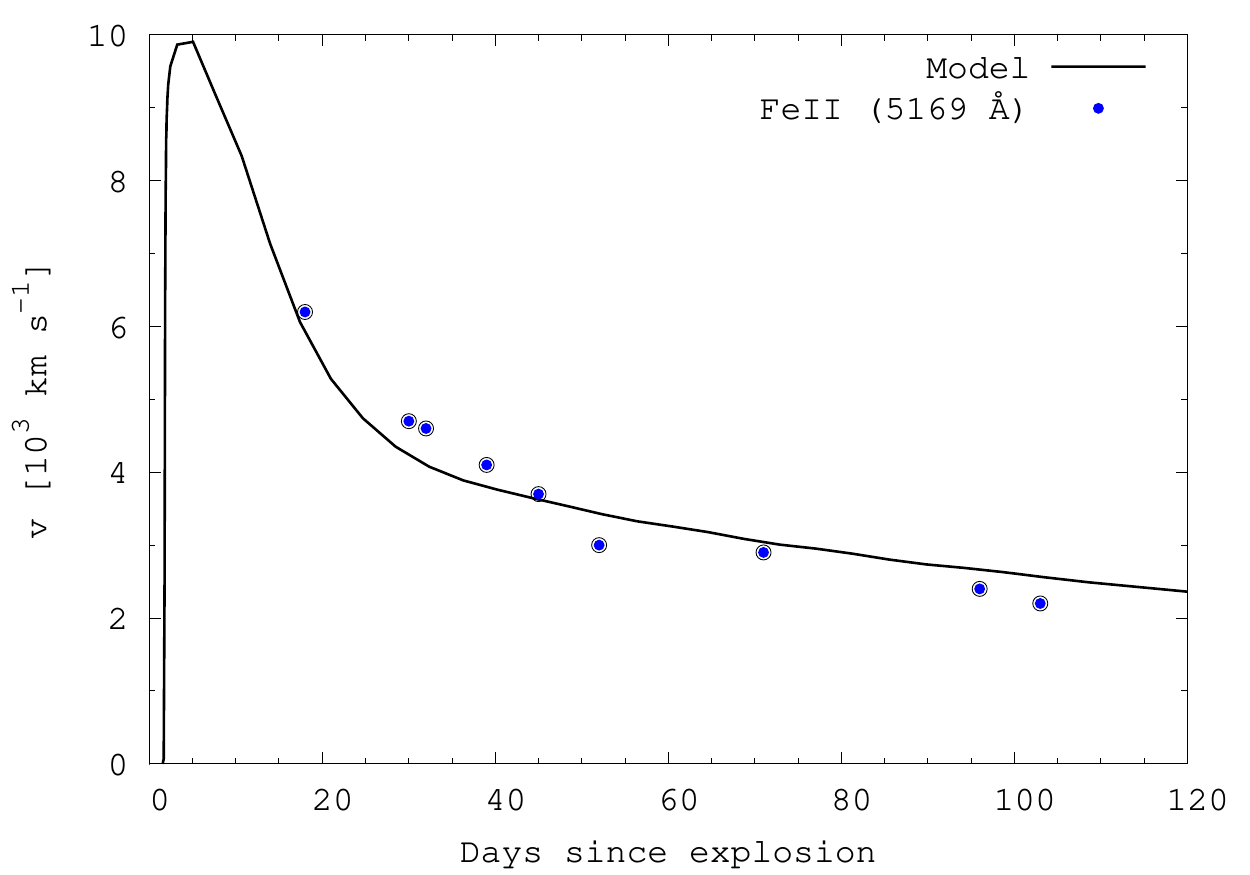}
        \begin{flushleft}
                \caption{Comparison between models and observations for SN 2012aw (top) and SN 2012ec (bottom). (Left) Bolometric light curves. (Right) Evolution of the photospheric velocity.}
                \label{fig:models2}
        \end{flushleft}
\end{figure*}

Our goal is to derive physical parameters ($M_{\rm hydro}$, $R$, $E$ and $M_{\rm Ni}$) for our sample (see Sect. \ref{sec:sample}) from the hydrodynamic modelling of LCs and photospheric velocities using the code described in Sect.\,\ref{sec:models}. As mentioned, we have decided to set the values of the radius of the progenitor to those derived from the analysis of pre-explosion images (see Table \ref{table:radios_preexp}).

We calculate a grid of models at fixed radius (inside the range of allowed values), varying $M_{\rm hydro}$, $E,$ and $M_{\rm Ni}$, to compare with observations. These parameters are modified until finding a model that best represents the observations. Our preferred model election is based on visual comparison. This procedure is extensively used in the literature despite the lack of statistical support. The main reason behind this is the typical lack of knowledge about the errors involved. A more robust method can be achieved by generating a grid of light curves in the parameter space and a quantified fitting procedure, as for example by $\mathbf{\chi^2}$ minimization. A further analysis of the confidence regions can determine the zones of degeneracy in parameter space and the effect on this degeneracy when considering the photospheric velocities. However, since it is difficult to accurately determine the uncertainties involved in the data and models, we decided not to perform a statistical analysis. Therefore, parameters derived by different authors may yield equally plausible solutions.

Our preferred models are presented in Figs. \ref{fig:models1} and \ref{fig:models2}, and in Table \ref{table:results} the physical parameters used for these models are shown. A range of validity for each parameter is also presented in Table \ref{table:results}. This range was found by making small changes to each optimal parameter and comparing with the observations (see discussion in Appendix \ref{ap:models_election} for more details). Therefore, these should not be interpreted as statistical uncertainties.

We find an overall good agreement between models and observations. The greatest differences appear during the earliest phase, known as adiabatic cooling. As noted in recent works \citep{yaron17,morozova18}, during this early stage the LC could be affected by the presence of circumstellar material (CSM). In particular, SN 2004et has been tested including a CSM material (see dashed line in model for SN 2004et, Fig. \ref{fig:models1}). It is clear that with the inclusion of CSM the comparison between model and observations improves considerably during early phases without changes at later times (Englert \& Bersten, in prep.). Some differences, although smaller, could be noticed during the transition between the plateau and the radioactive tail. During this phase, the object is almost completely recombined and the notion of a photosphere loses meaning. Therefore during this stage our models begin to be less reliable.

\begin{table}
\caption{Physical parameters derived from the hydrodynamic modelling. A range of validity for each parameter is also presented. We note that these are not statistical errors (see discussion in Sect.\,\ref{sec:results} and Appendix \ref{ap:models_election}).}
\label{table:results}   
\resizebox{0.5\textwidth}{!}{
\centering                
\begin{tabular}{c c c c c}      
        \hline\hline\noalign{\smallskip}    
        SN & M$_{\rm hydro}$ & Radius & Energy & M$_{\rm Ni}$ \\
        & $[\rm M_{\odot}]$ & $[\rm R_{\odot}]$ & [foe]  & $[\rm M_{\odot}]$ \\
        \hline\noalign{\smallskip}       
        2004A  & 10 $\pm$ 0.5 & 1000$^{+300}_{-100}$ & 0.45$^{+0.05}_{-0.03}$ & 0.085 $\pm$ 0.005 \\
        \noalign{\smallskip}  
        2004et & 18 $\pm$ 1.0 & 1250$^{+200}_{-100}$ & 1.2 $\pm$ 0.1  & 0.063 $\pm$ 0.005 \\
        \noalign{\smallskip}  
        2005cs & 12 $\pm$ 1.0 & 400 $\pm$ 50 & 0.33 $\pm$ 0.05 & 0.0015 $\pm$ 0.0002 \\
        \noalign{\smallskip}  
        2008bk & 11$^{+0.5}_{-1.0}$ & 650$^{+150}_{-50}$  & 0.2$^{+0.03}_{-0.02}$ & 0.0085 $\pm$ 0.0005 \\
        \noalign{\smallskip} 
        2012aw & 23$^{+1}_{-2}$ & 800 $\pm$ 100 & 1.4 $\pm$ 0.2 & 0.066 $\pm$ 0.006 \\
        \noalign{\smallskip} 
        2012ec & 10 $\pm$ 1 & 1000 $\pm$ 150 & 0.6 $\pm$ 0.05 & 0.042 $\pm$ 0.003 \\
        \hline
\end{tabular}}
\end{table}

Although our goal was to set the radius of the progenitor within the values determined by the pre-explosion detections, there were two cases, SNe 2004A and 2004et, for which this was not possible. Larger values were needed to model their LCs, since with the values found in the literature, all models indicated much lower plateau luminosities than those observed. The largest discrepancy is found for SN 2004et. For this SN, the progenitor identification is not entirely clear (see Sect.\,\ref{subsec:04et}) which may introduce an error in the derived radius.

The analysis of the whole sample implies the following range for the physical parameters: $M_{\rm hydro}$ = 10 -- 23\,$M_{\odot}$, $R$ = 400 -- 1250\,$R_{\odot}$, $E$ = 0.2 -- 1.4\,foe (1\,foe $\equiv$ 10$^{51}$erg) and $M_{\rm Ni}$ = 0.0015 -- 0.085\,$M_{\odot}$, as Table \ref{table:results} shows. It is interesting that despite the fact that the sample is small (only six objects), we find a wide range in the explosion parameters. This seems to indicate that there is a great diversity in the properties of SNe II-P progenitors.

\section{Analysis}
\label{sec:analysis}
\subsection{Correlations between physical parameters}

In the previous section, we derived the physical parameters that characterise the SN explosion, namely, the mass and radius of the progenitor before explosion, the energy released during the core collapse, and the amount of radioactive material synthesised in the explosion. Here, we analyse possible correlations between different parameters.

\begin{figure}
\flushleft
        \includegraphics[width=0.5\textwidth]{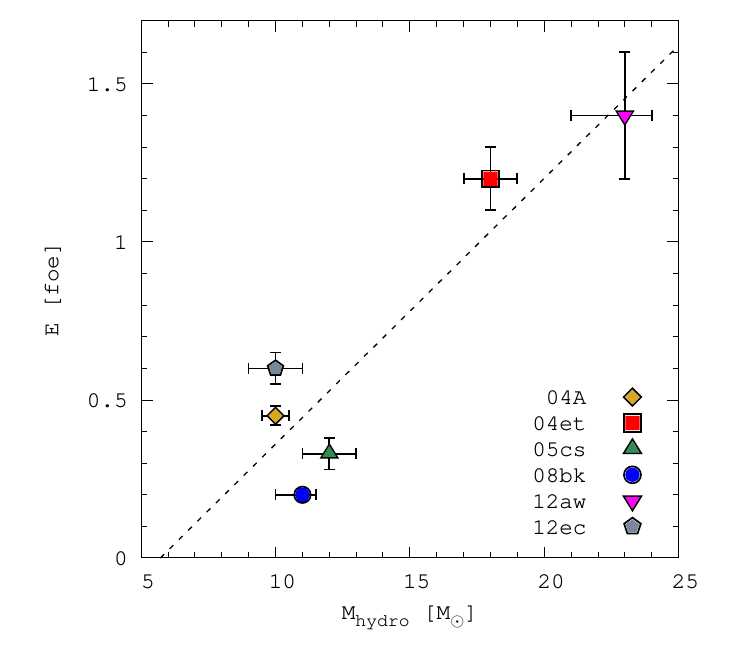} \\
        \begin{flushleft}
                \caption{Analysis of possible correlations between different physical parameters. In this case we present explosion energy as a function of $M_{\rm hydro}$. Dashed line shows a linear fit to data.}
                \label{fig:corr_masa_energia}
        \end{flushleft}
\end{figure}

In Fig. \ref{fig:corr_masa_energia} we present the relation between $M_{\rm hydro}$ and $E$. From the figure it is clear that these parameters seem to be correlated, in the sense that more-massive objects seem to generate higher-energy explosions in agreement with previous studies in the literature \citep{utrobin15,pejcha15}. In the same figure we also present a linear regression to the data. We implemented a Pearson's chi-squared test to analyse how significant the correlation is. We found a value of $\rho = 0.91$ for the mass-energy correlation, which confirms the strong correlation between them.

\begin{figure}
\centering
        \includegraphics[width=0.5\textwidth]{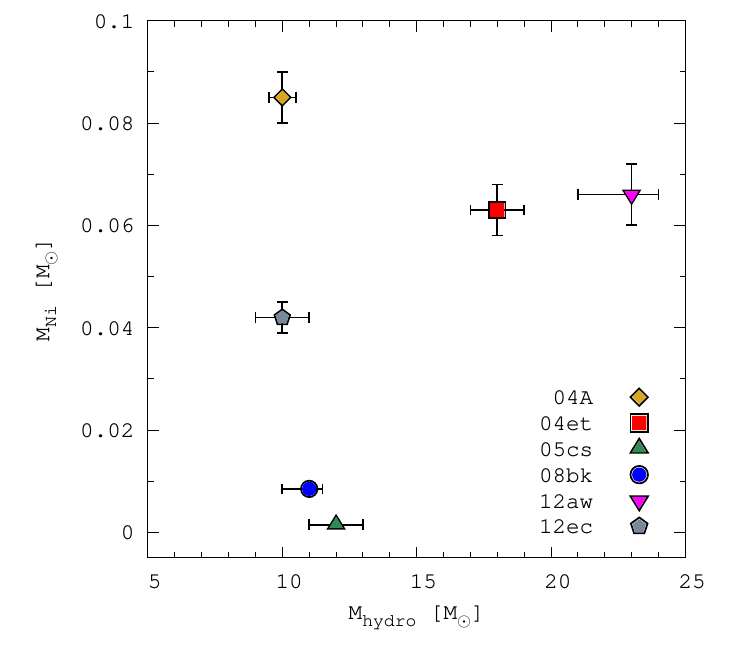} \\
        \begin{flushleft}
                \caption{Analysis of possible correlations between different physical parameters. We present $M_{\rm Ni}$ as a function of $M_{\rm hydro}$.}
                \label{fig:corr_masa_niquel}
        \end{flushleft}
\end{figure}

\begin{figure}
\centering
        \includegraphics[width=0.5\textwidth]{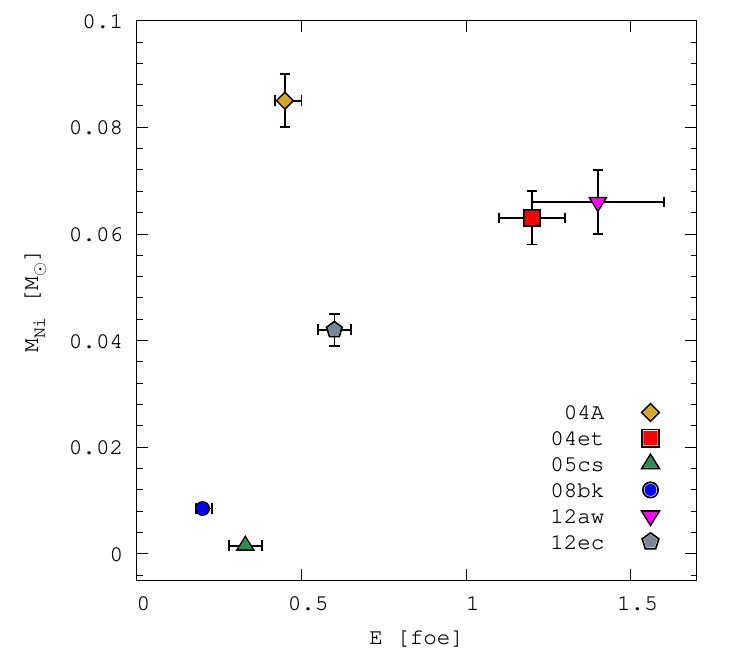}
        \begin{flushleft}
                \caption{Analysis of possible correlations between different physical parameters. In this case we present $M_{\rm Ni}$ as a function of explosion energy.}
                \label{fig:corr_energia_niquel}
        \end{flushleft}
\end{figure}

Figure \ref{fig:corr_masa_niquel} shows derived values of $M_{\rm hydro}$ and $M_{\rm Ni}$. Again, there seems to be a tendency between both parameters, in the sense that more-massive objects seem to produce more radioactive material. However, there is a significant dispersion in the relationship which is reflected in the value of the Pearson coefficient with $\rho = 0.34$. The same tendency is observed when analysing a possible correlation between the explosion energy and $M_{\rm Ni}$ (see Fig. \ref{fig:corr_energia_niquel}), suggesting that explosions that release more energy seem to produce a greater amount of radioactive material as predicted by modelling of the explosive nucleosynthesis \citep{woosley95}. On this occasion, the correlation coefficient gave us a value of $\rho = 0.6$ pointing out the existence of a tendency. We note that SN 2004A could be the one responsible for deviations in these trends due to its large amount of \Ni\, and its low mass and explosion energy. Such a large amount of \Ni\, is necessary to reproduce the luminosity of the SN in the radioactive tail phase, which is an almost direct indicator of the amount of nickel produced. However, this value strongly depends on the distance assumed and also, although to a lesser extent, on the explosion epoch used. In addition, the value of the explosion time also affects the estimations on the mass and explosion energy. As discussed in Sect.\,\ref{subsec:04a}, the value of $t_{\rm exp}$ for SN 2004A is not well constrained. We also note that as opposed to the rest of the SNe in our sample, this is the only object which clearly was not observed during the cooling phase (see Figs. \ref{fig:sample_lc} and \ref{fig:models1}). Therefore, we believe that the poor estimation of the explosion time may be responsible for the deviations on the relation observed in Figs. \ref{fig:corr_masa_niquel} and \ref{fig:corr_energia_niquel}. Excluding this SN from this analysis, the value of the Pearson coefficent is $\rho = 0.76$ and $\rho = 0.95$ for the relations $M_{\rm hydro}$ -- $M_{\rm Ni}$ and $E$ -- $M_{\rm Ni}$, respectively, suggesting a clear correlation.

We studied relations between other parameters, but these are not presented here since no significant correlation was found. 

The observed correlations seem to indicate that more-massive stars release more energy during core collapse, and therefore they can synthesise more radioactive material. Although our sample is too small to reach a definitive conclusion, our results agree with those of \cite{utrobin15}.

\subsection{Comparison with previous physical--observed parameter relations}

The progenitor and explosion properties of SNe II-P can be studied in a number of ways. We have already mentioned the hydrodynamic modelling of their LCs and the analysis of direct detections in pre-explosion images. Additionally, more than three decades ago, it was proposed that certain observables of SNe can be used to determine their physical parameters. The observational properties of a SN such as the plateau luminosity ($L_{p}$), expansion velocity ($v_{exp}$), and plateau length ($\Delta t_{p}$) can be measured, and models can then be used to determine the explosion parameters such as the ejected mass ($M_{ej}$), explosion energy ($E$), and pre-SN radius of the star ($R$).

The relations between physical parameters and observables were first derived analytically by \cite{arnett80} and then generalised by \cite{popov93}. Numerical calibrations of these relations were then given by \citet[][hereafter LN83 and LN85, respectively]{litvinova83,litvinova85} based on a grid of hydrodynamic models for different values of $M_{ej}$, $R$, and $E$. These calibrations use observational properties at mid-plateau ($L_{p}$, $v_{exp}$) and plateau length as input to estimate the physical parameters. They are widely used in the literature since with simple measurements of observed parameters, physical parameters of the progenitors can be derived. However, conclusions obtained with this method need to be analysed carefully since LN83 used simplified models. In particular, \cite{hamuy01} used these relations to derive parameters from a sample of 16 objects, obtaining in some cases unrealistic parameters.

\cite{kasen09} presented updated models through the calculation of LCs and spectra for different masses, metallicities, and explosion energies, using initial models coming from stellar evolution calculations. These latter authors use their models to describe the dependence of plateau luminosity and duration on explosion energy and progenitor mass. Nevertheless, the relations they found are only simple and easy to apply in the extreme case of no $^{56}$Ni production. When $^{56}$Ni is considered, the relations involve more parameters, which complicate their application to obtain physical parameters from observations. In addition, the mass and radius are not treated as independent values because they used stellar-evolution models as initial configuration. Therefore, we cannot directly compare our results using these relations. 

Here we compare our results with those obtained using the LN85 relations. For this purpose, we measured the aforementioned parameters: $M_{V}$, $v_{exp}$, and $\Delta t_{p}$. When we did not have measurements of magnitudes or velocities right in the middle of the plateau, we implemented a linear interpolation to obtain them. Using those observables and the LN85 relations we derived physical properties of each SN of our sample. These results are presented in Table \ref{table:parameters_ln}.

\begin{table}
\centering
        \caption{Physical parameters of the sample using LN85 relations.}
        \label{table:parameters_ln}
        \medskip
        \begin{tabular}{ccccccc}
                \hline\hline\noalign{\smallskip}
                SN & $\Delta t_{p}$ & $v_{exp}$ & $M_{V}$ & $M_{ej}$ & $R$ & $E$ \\
                   & [days]     & [km s$^{-1}$] &  & [$M_{\odot}$] & [$R_{\odot}$] & [foe]\\
                \hline\noalign{\smallskip}
                2004A  & 113 & 3078 & -16.6 & 16 & 407 & 0.8 \\
                2004et & 120 & 3605 & -17.0 & 22 & 420 & 1.3 \\
                2005cs & 119 & 1600 & -14.6 & 16 & 166 & 0.2 \\
                2008bk & 131 & 2034 & -14.8 & 28 & 115 & 0.5 \\
                2012aw & 128 & 3500 & -16.6 & 30 & 250 & 1.5 \\
                2012ec & 100 & 3546 & -16.6 & 15 & 315 & 0.9 \\
                \hline
        \end{tabular}
\end{table}

\begin{figure}
\centering
        \includegraphics[width=0.5\textwidth]{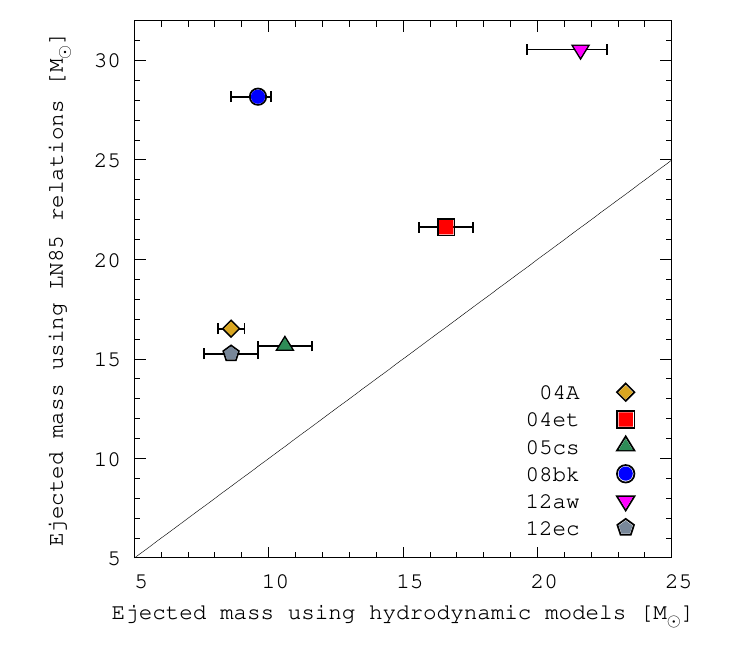}
        \begin{flushleft}
        \caption{Ejected masses obtained with the LN85 relations compared to those obtained by hydrodynamical modelling of LCs and photospheric velocities.}
        \label{fig:ln85_masses}
        \end{flushleft}
\end{figure}

\begin{figure}
\centering
        \includegraphics[width=0.5\textwidth]{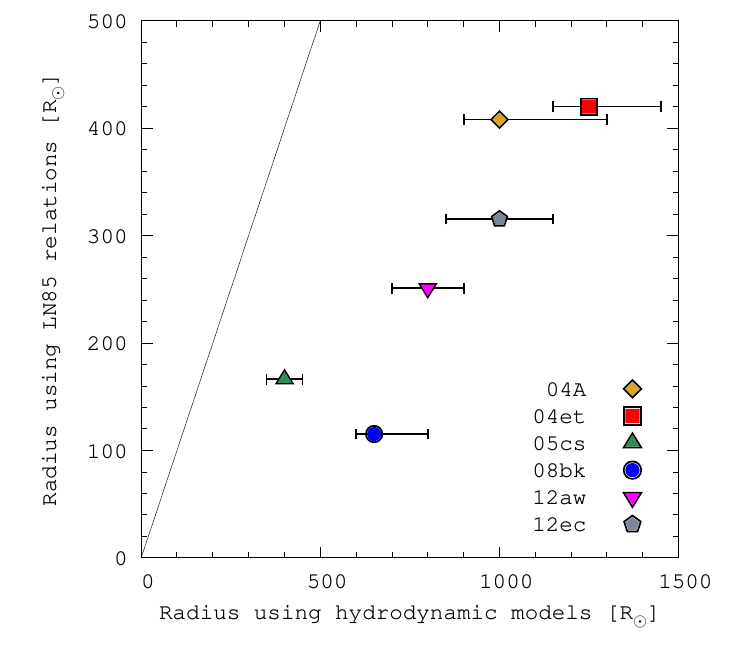}
\begin{flushleft}
        \caption{Progenitor radii obtained with the LN85 relations compared to those obtained by our analysis.}
        \label{fig:ln85_radii}
\end{flushleft}
\end{figure}

\begin{figure}
\centering
        \includegraphics[width=0.5\textwidth]{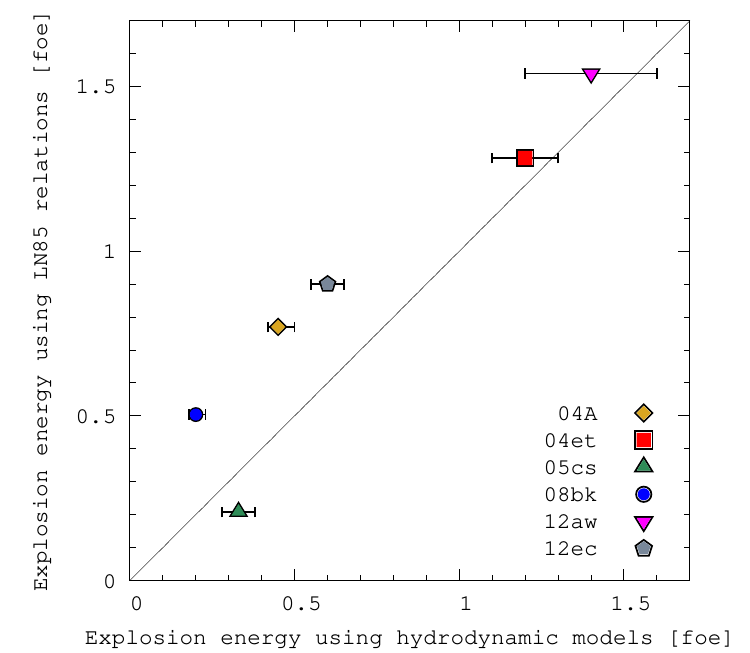}
\begin{flushleft}
        \caption{Explosion energy obtained with the LN85 relations compared with those obtained by our modelling.} 
        \label{fig:ln85_energies}
\end{flushleft}
\end{figure}

Figures \ref{fig:ln85_masses} to \ref{fig:ln85_energies}  show the comparison of our results for masses, radii, and explosion energies, respectively, with those obtained through the LN85 relations. From Fig. \ref{fig:ln85_masses} it is clearly seen that the ejected masses calculated using the LN85 relations are systematically larger than ours. We note that the ejecta mass in our models is derived by subtracting the mass of the compact remnant, which was considered to be of 1.4\,$M_{\odot}$, to the hydrodynamical mass presented in Table \ref{table:results}. The smallest differences between both results are already too large, being around 5\,$M_{\odot}$ (for SNe 2004et and 2005cs) while the largest difference appears for SN 2008bk, being of the order of 18\,$M_{\odot}$. Taking into account all the SNe in the sample, we obtain an  average separation of 8.5\,$M_{\odot}$ when comparing our results with those from LN85 relations.
From Fig. \ref{fig:ln85_radii} we can see that radii estimated with LN85 relations are substantially and systematically lower than ours. It must be taken into account that the radius that we used was derived from the analysis of direct detections on pre-explosion images, except for SNe 2004A and 2004et for which this was not possible (see discussion in Sect. \ref{sec:models}). On average, the separation between both results is 570\,$R_{\odot}$ while our results are $\sim$\,3.3 times larger. Figure \ref{fig:ln85_energies} shows the relation between explosion energies. We note that in most cases, the explosion energy found using the LN85 relations also provides larger values than ours, but the differences here are smaller than for the other parameters.

This analysis shows that there are significant differences in the parameters derived by both methods. This is likely because the LN83 models do not include the effect of heating due to radioactive decay; they use old opacity tables without considering any opacity floor, and simplified pre-SN models, such as single polytropic models that do not reproduce the inner part of the progenitor. In addition, the methodology used to derive the physical properties in these two works is different. We model the complete LC together with the evolution of the photospheric velocities, while LN85 uses only three observables to derive the progenitor and explosion properties. Even though connecting observables with physical parameters using simple relations could be very useful for large data sets, it is clear that these relations do not seem to give reliable results.

\subsection{Comparison with previous works using hydrodynamical modelling}
\label{sec:comparison_hydro}

We compare our results with previous hydrodynamical modelling of the same SNe available in the literature. These are the works of \citet{utrobin08, utrobin09} on SNe 2004et and 2005cs, \citet{morozova18} on SNe 2004et, 2005cs, 2012aw, and 2012ec, \citet{pumo17} on SNe 2005cs, 2008bk, and 2012aw, and \citet{ricks19} on SN 2004et. Table \ref{table:comparacion} summarises the ejecta masses found by these latter authors compared with ours. As can be seen, the results from Utrobin \& Chugai show values larger than ours and those from \citet{morozova18} and \citet{ricks19} are commonly lower, while our results agree quite well with those of \citet{pumo17}.

\citet{pumo17} use a semi-analytic code that solves the energy balance equation for ejecta of constant density in homologous expansion, and a general-relativistic, radiation-hydrodynamics Lagrangian code that simulates the evolution of the physical properties of the ejecta. They arrive at their best models by simultaneously fitting the LC, the continuum temperature, and the velocity evolution using a $\chi^2$ minimization method.

Although \citet{morozova18} use a hydrodynamic code similar to the one used in this work and a detailed analysis of the confidence regions in parameter space to reach their preferred models, they only use the LC as an observable in their fitting process, thus ignoring important constraints introduced by the photospheric velocities. We computed similar progenitors to those of \citet{morozova18} using the public stellar evolution code \texttt{MESAstar\footnote{\url{http://mesa.sourceforge.net/}}} version 10398 \citep{paxton11,paxton13,paxton15,paxton18} and got generally good agreements in the LCs, except for SN 2004et. However, in all cases the photospheric velocities obtained in the models underestimate the velocities of the SNe. The discrepancies in progenitor masses (see Table \ref{table:comparacion}) may be due to the lack of velocity fits. Recently, \citet{ricks19} modelled SN 2004et using the hydrodynamic code STELLA to simulate LCs, and photospheric and Fe\,{\sc ii} velocities. They found a similar progenitor mass to that of \citet{morozova18} but a higher explosion energy.
 
A more sophisticated code is used in the works of Utrobin \& Chugai, both with respect to the radiative transfer and to the treatment of the matter, which takes into account non-LTE effects on the average opacities and the thermal emissivity, effects of non-thermal ionisation, and a contribution of lines to the opacity, among other effects. We computed similar progenitors to those adopted in the aforementioned works, and obtained comparable LCs. This is to emphasise the existence of degeneracy in the parameter space that allows  almost identical LCs to be obtained for more than one set of physical parameters. As our work was being completed, \citet{dessart19} and \citet{goldberg19} submitted two papers emphasising that light curve modelling cannot provide a unique solution for the ejecta mass of SNe II.

\begin{table}
\centering
        \caption{Derived ejected masses in the literature for the SNe in our sample.}
        \label{table:comparacion}
        \medskip
        \begin{tabular}{cccccc}
                \hline\hline\noalign{\smallskip}
                 & 2004et & 2005cs & 2008bk & 2012aw & 2012ec \\
                \hline\noalign{\smallskip}
                This work & 16.6 & 10.6 & 9.6  & 21.6 & 8.6 \\
                1                 & 12.5 & 7.8  &  -   & 14.0 & 8.7 \\
                2                 &  -   & 9.5  & 10.0 & 19.6 &  -  \\
                3, 4      & 22.9 & 15.9 &  -   &  -   &  - \\
                5		  & 11.9 &	-	&	-	&	-	&	- \\
                \hline
        \end{tabular}
        \tablebib{(1) \citet{morozova18}; (2) \citet{pumo17}; \linebreak (3) \citet{utrobin08}; (4) \citet{utrobin09}; (5) \citet{ricks19}.}
\end{table}

\subsection{Comparison with results obtained from the analysis of pre-explosion images}
\label{sec:comp_pre_exp}

One of the main goals of this work is to test whether or not systematic differences between the masses derived using pre-explosion observations and hydrodynamic models exist, as suggested by previous studies \citep{utrobin08}.

As we have mentioned before, the mass determined by our hydrodynamic models corresponds to the mass of the star just before the explosion, and therefore it is usually smaller than the mass of the star in the main sequence ($M_{\rm hydro} \lesssim$ $M_{\rm ZAMS}$) due to mass loss during evolution. On the other hand, the mass derived from the pre-explosion images is the mass of the star in the main sequence ($M_{\rm ZAMS}$), since it is derived by connecting an evolutionary track with the position of the star in the HR diagram.

Therefore, to determine whether or not the masses are compatible,  we must analyse the magnitude of the difference between the initial and final mass, that is, the amount of mass lost during the evolution. We use the stellar evolution code \texttt{MESA} to obtain the pre-SN mass ($M_{\rm preSN}$) for those stars with initial masses in the ranges shown in Table \ref{table:radios_preexp}. In cases where there is more than one value of the mass derived from pre-explosion images, we use ranges of values that include all those values of $M_{\rm ZAMS}$ with their respective errors, derived by each author. We evolve stars from the pre-main sequence assuming an initial metallicity of $Z = 0.02$. For every model, we use the “Dutch” wind scheme defined in the \texttt{MESA} code \citep{dejager88,vink01,glebbeek09}.

Mass loss is a critical phenomenon in massive-star evolution and is one of the channels by which massive stars affect their environment. Despite its importance, our knowledge about mass loss is not complete. One of the processes by which massive stars lose material is by radiative winds driven via lines or dust. In line-driven winds, momentum is transferred from photons to the gas via absorption and line scattering. The presence of inhomogeneities (e.g. clumps) in the stellar atmospheres can complicate the situation, as these introduce changes in the derived mass-loss rates. It is now well established that winds are clumpy \citep[see, e.g.][]{evans04,bouret05,fullerton06}, and therefore mass-loss rates can be overestimated when homogeneous winds are assumed. Recent works suggests that the algorithms used in stellar-evolution calculations may yield mass-loss rates that are too high by  a factor of between two and ten. Particularly, \citet{puls08} and \citet{smith14} suggest that the algorithms are overestimated by a factor of three \citep[see also][for details]{renzo17}. Therefore we use two different values for the wind efficiency, $\eta = 1.0$ and 0.33, in the calculations of final masses.

\begin{figure}
\centering
        \includegraphics[width=0.45\textwidth]{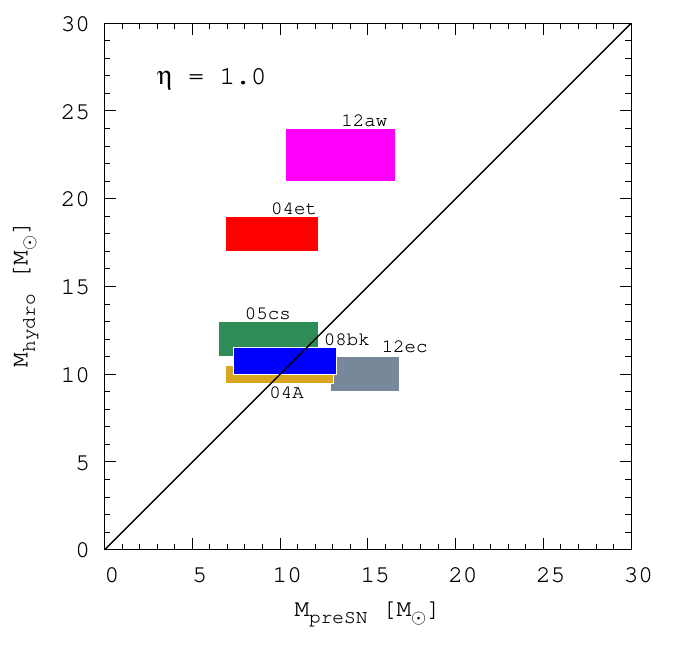}
        \includegraphics[width=0.45\textwidth]{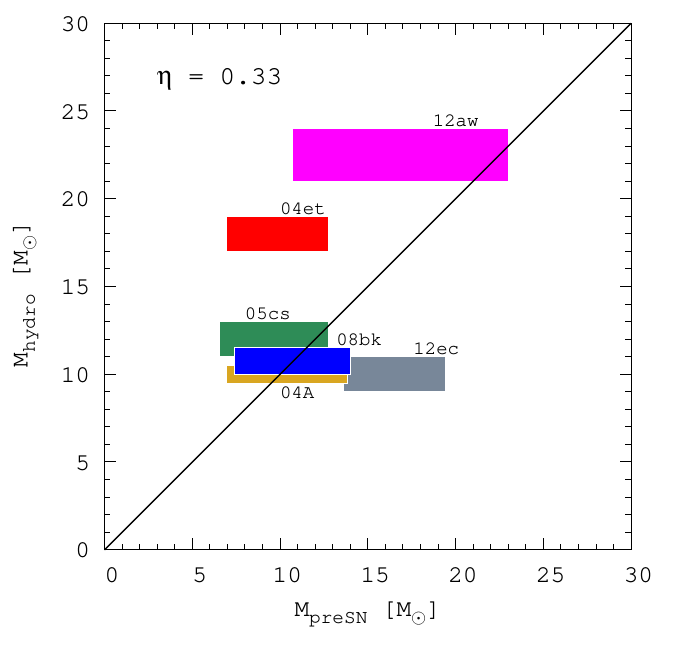}
\begin{flushleft}
        \caption{Comparison between the masses obtained in this work using hydrodynamic models ($M_{\rm hydro}$) and the final masses of the progenitors ($M_{\rm preSN}$) determined using a stellar evolution code. The panels correspond to different values for the wind efficiency: $\eta = 1.0$ (upper) and $\eta = 0.33$ (bottom).}
        \label{fig:comparison_masses_wind}
\end{flushleft}
\end{figure}

Figure \ref{fig:comparison_masses_wind} compares our hydrodynamic masses and pre-SN masses for the different wind efficiencies. When using $\eta = 1.0$ (upper panel) we can see that for SNe 2004et and 2012aw, our hydrodynamical mass overestimates the pre-explosion mass. Using $\eta = 0.33$ (bottom panel), we notice that SN 2004et is the only one that overestimates the pre-explosion mass. It is interesting to note that the identification of the progenitor for SN 2004et is not entirely clear; it could therefore be that the pre-SN properties derived from these images are not entirely correct (see discussion in Sect. \ref{subsec:04et}).

From this analysis we conclude that unlike what was found in previous works, for example different results from \cite{utrobin07}, \cite{utrobin08,utrobin09,utrobin13,utrobin15,utrobin17}, Fig. 8 of \cite{morozova18}, or Fig. 25 of \cite{maguire10}, 
we find good agreement between the masses estimated by hydrodynamic models and those obtained by the analysis of pre-explosion images. Despite the small size of the sample studied here, all our objects have secured progenitor identifications and well-sampled photometric and spectroscopic monitoring allowing proper modelling of the LC and velocity evolution. There are currently no more data available to make a rigorous comparison between both methods.

\section{Conclusions} 
\label{sec:conclusion}

We derived physical properties through hydrodynamical modelling for a sample of well-observed SNe, namely SN 2004A, 2004et, 2005cs, 2008bk, 2012aw, and 2012ec. These are all SNe II-P for which there exists enough photometric and spectroscopic monitoring during the plateau and radioactive phases to allow reliable hydrodynamical modelling, pre-explosion images with direct information from the progenitor star, and post-explosion images confirming the disappearance of the progenitor. A short version of this study was presented in \citet{martinez18}. Analysing the LCs of our sample, we note that there is a large amount of variation in the luminosities of the plateau and radioactive tail, and in the plateau length. The following ranges of physical parameters were estimated for the whole sample: $M_{\rm hydro}$ = 10 -- 23\,$M_{\odot}$, \linebreak R = 400 -- 1250\,$R_{\odot}$, E = 0.2 -- 1.4\,foe and $M_{\rm Ni}$ = 0.0015 -- 0.085\,$M_{\odot}$. The wide range of parameters found even for a small sample of six SNe is consistent with the variety of observed properties among the objects. Interestingly, SNe 2005cs and 2008bk show similar velocity and luminosity evolution, except during the radioactive phase, where SN 2008bk is substantially more luminous. This seems to indicate that while both objects share similar progenitor properties, they experienced very different  nucleosynthesis, or part of this synthesised material has remained within the compact remnant.

Due to the existence of a degeneracy between mass, radius, and explosion energy, we chose to model our LCs and photospheric velocities adopting the progenitor radius value from the direct detections. However, for two objects (SNe 2004A and 2004et) this was not possible; we had to assume larger radii than those derived in the literature in order to match the plateau luminosities. The largest discrepancy is found for SN 2004et. For this SN, the progenitor identification is not entirely clear (see Sect.\,\ref{subsec:04et}), which may introduce an error in the derived radius.

From the analysis of the sample we searched for correlations between different physical properties. We conclude that a strong correlation is found between progenitor mass and explosion energy, in the sense that more-massive objects seem to cause higher-energy explosions, as found in previous studies \citep{utrobin15}. When analysing possible correlations that involve $M_{\rm Ni}$ with pre-SN mass and explosion energy, we infer that there seems to be a tendency in the sense that more-massive objects, and therefore higher-energy explosions, produce larger amount of radioactive material, although the dispersion in these relations is large. In both cases, SN 2004A shows the largest deviation from these trends due to its larger amount of \Ni\, for its pre-SN mass and explosion energy. We think that this deviation is likely due to an incorrect determination of the explosion date, which in turn affects the estimation of progenitor properties.

We also compare our results with those obtained using the LN85 relations. We found  significant differences between the parameters derived using the two methods. On average, ejected mass estimations using the LN85 relations are $\sim$1.75 times larger than ours, while our estimation of pre-SN radius is $\sim$3.3 larger than that of LN85. These differences could be due to the facts that the LN83 models do not include the effect of radioactive heating, they use old opacity tables, and they adopt simplified pre-SN models. In addition, LN85 use only three observables to derive those physical parameters whereas we modelled the complete LC and the photospheric velocities, and have fixed the radius value, which allowed us to reduce the number of free parameters of the model. Even though the idea of having simple relations to connect observables with physical parameters could be very useful to apply to large datasets, due to the limitations and simplifications of these models and relations, we believe that our results are more reliable and caution should be exercised when using these relations.

Finally, we compare the masses we obtained using hydrodynamic models with those that have been determined from direct detections in pre-explosion images. We find that our determination of progenitor mass is not systematically larger, as was found in the literature \citep[see discusion in][]{utrobin08,maguire10,morozova18}. This shows that in some cases the two methods for determining physical properties of progenitors give consistent results. Perhaps, the differences found in the literature are due to the simplified models used to derive those parameters or to the use of objects whose progenitor candidates are not confirmed. Indeed, we note that using similar pre-SN models with the same sets of parameters as \citet{utrobin08,utrobin09}, we arrive at similar LCs. This demonstrates the high degree of degeneracy present in this problem even when modelling LCs together with the photospheric velocity evolution. Nevertheless, we emphasise the fact that we were able to find a set of parameters in accordance with the radius determined by pre-explosion images and with a progenitor mass compatible with the one estimated by direct detections, one of the major aims of this work.

\begin{acknowledgements}

We want to thank Giuliano Pignata for providing photometry of SN 2008bk. We are also grateful to Schuyler Van Dyk for his help in the confirmation of the progenitor of SN\,2012ec. Special thanks to Gastón Folatelli for his generous revision of the manuscript which helped clarify and improve the paper.

\end{acknowledgements}

\textit{Software:} \texttt{MESA} \citep{paxton11,paxton13,paxton15,paxton18}, \texttt{gnuplot} \citep{williams15}


\begin{appendix}
\section{Pre-supernova models}
\label{ap:stellar_evolution}

\begin{figure*}
\centering
                \includegraphics[width=0.48\textwidth]{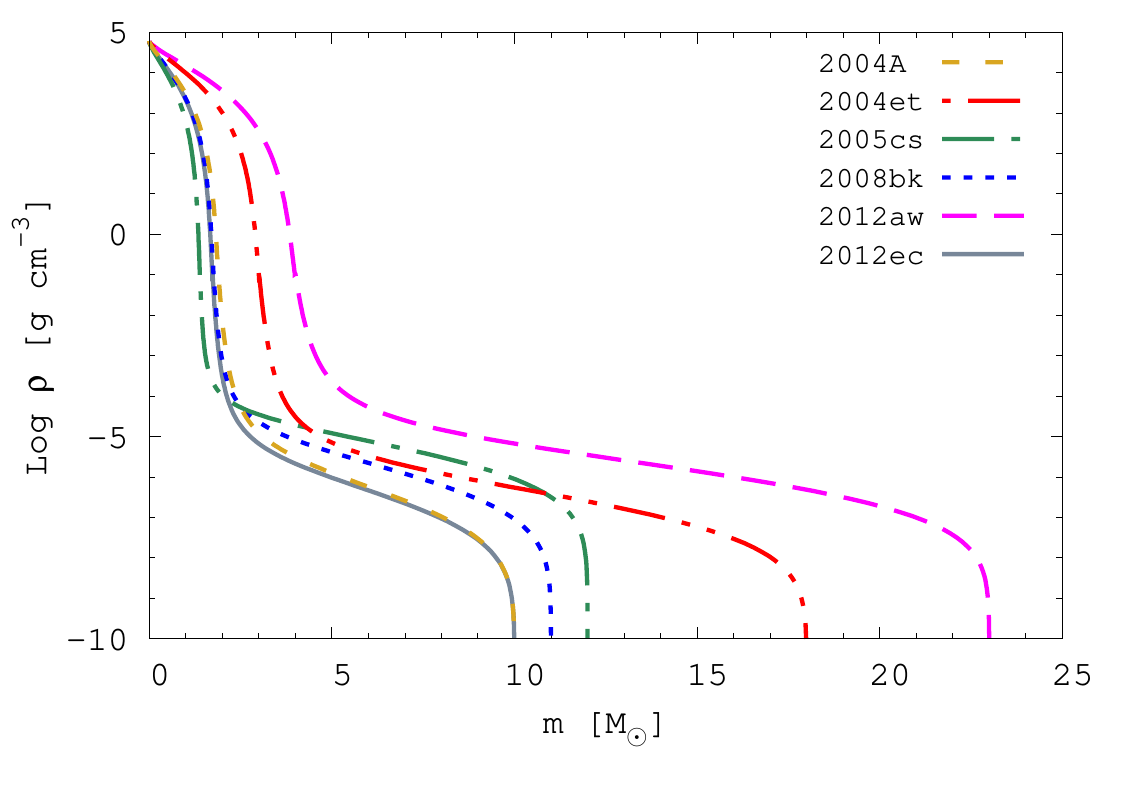} 
                \includegraphics[width=0.48\textwidth]{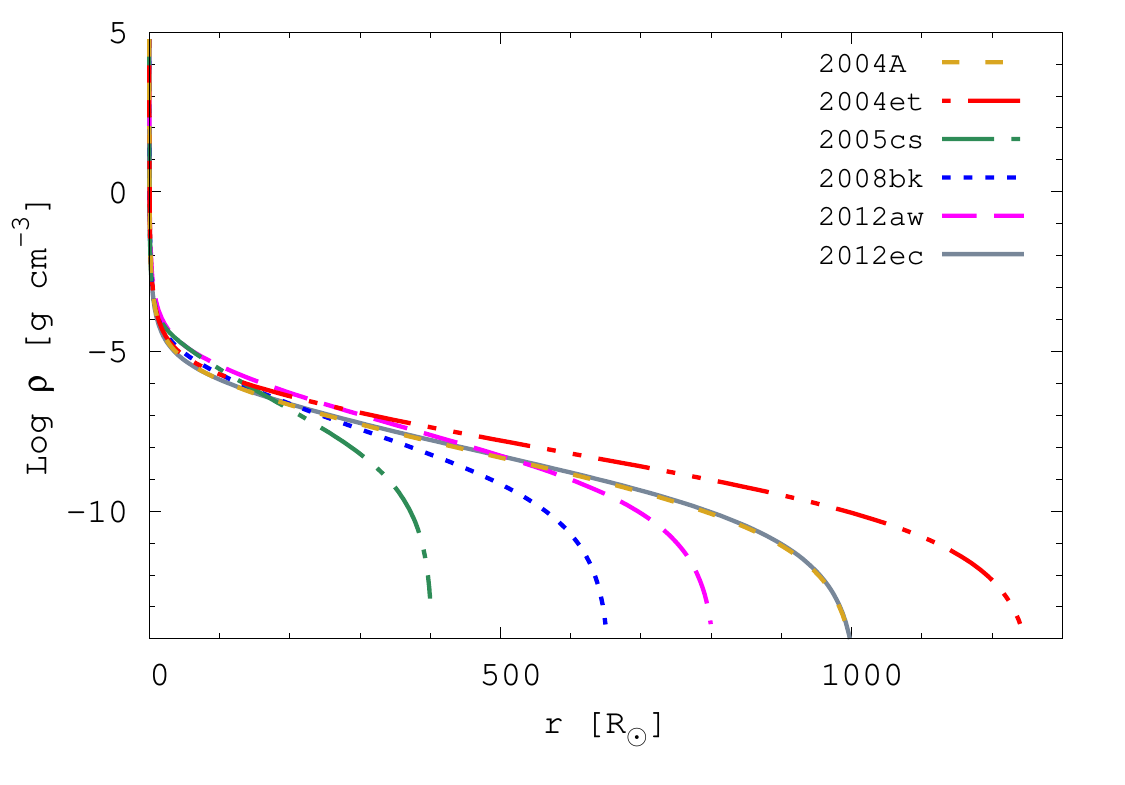} 
        \begin{flushleft}
                \caption{\emph{Left:} Pre-SN model for each SN of our sample as a function of mass. \emph{Right:} Pre-SN model for each SN of our sample as a function of radius. These models come from double polytropic calculations (see Sect. \ref{sec:models}). The values of the parameters for each model are shown in Table \ref{table:results}.}
                \label{fig:dens_profile}
        \end{flushleft}
\end{figure*}

Here, we complete the description of the pre-SN models used in our sample. Figure \ref{fig:dens_profile} shows the initial density profile used for each SN as a function of mass and radius. The initial structure is composed of a dense core and an extended envelope, which is characteristic of a RSG star.

There is a relationship between the mass of the helium core and the mass of the star in the ZAMS. That is, at first order, more-massive stars develop more-massive helium cores. This can be seen in any stellar evolution calculation that follows the evolution until the core collapse assuming that no other physical ingredients are taken into account, such as different treatment of overshooting, rotation, and so on \citep[see, e.g.][]{woosley02}. Therefore, objects which show a transition between the dense core and the envelope at higher mass values approximately correspond to more-massive objects in the ZAMS. In the same way, objects with the same core mass correspond to objects with similar $M_{\rm ZAMS}$. Therefore, it is possible to have two configurations with the same final pre-SN mass that may correspond to different masses in the main sequence. Thus, in Fig. \ref{fig:dens_profile}, we note that the cores of progenitor stars of SNe 2004A, 2005cs, 2008bk, and 2012ec are very similar, which would suggest similar masses in the ZAMS, while SNe 2004et and 2012aw seem to come from more-massive objects. This statement must be taken with caution however, as there may be other factors that affect the structure of the star at the moment of explosion. 

\begin{figure*}
\centering
                \includegraphics[width=0.46\textwidth]{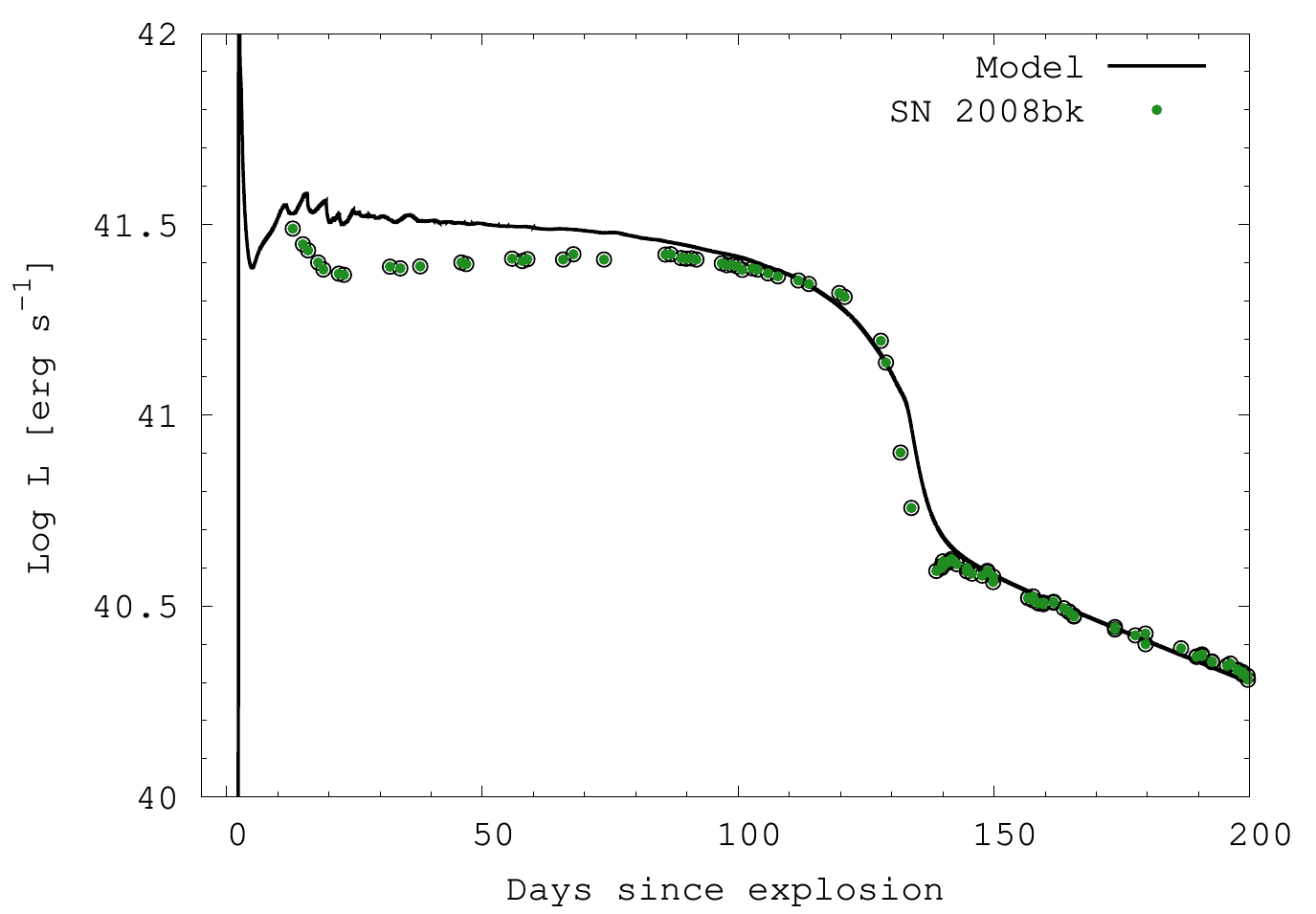}
                \includegraphics[width=0.46\textwidth]{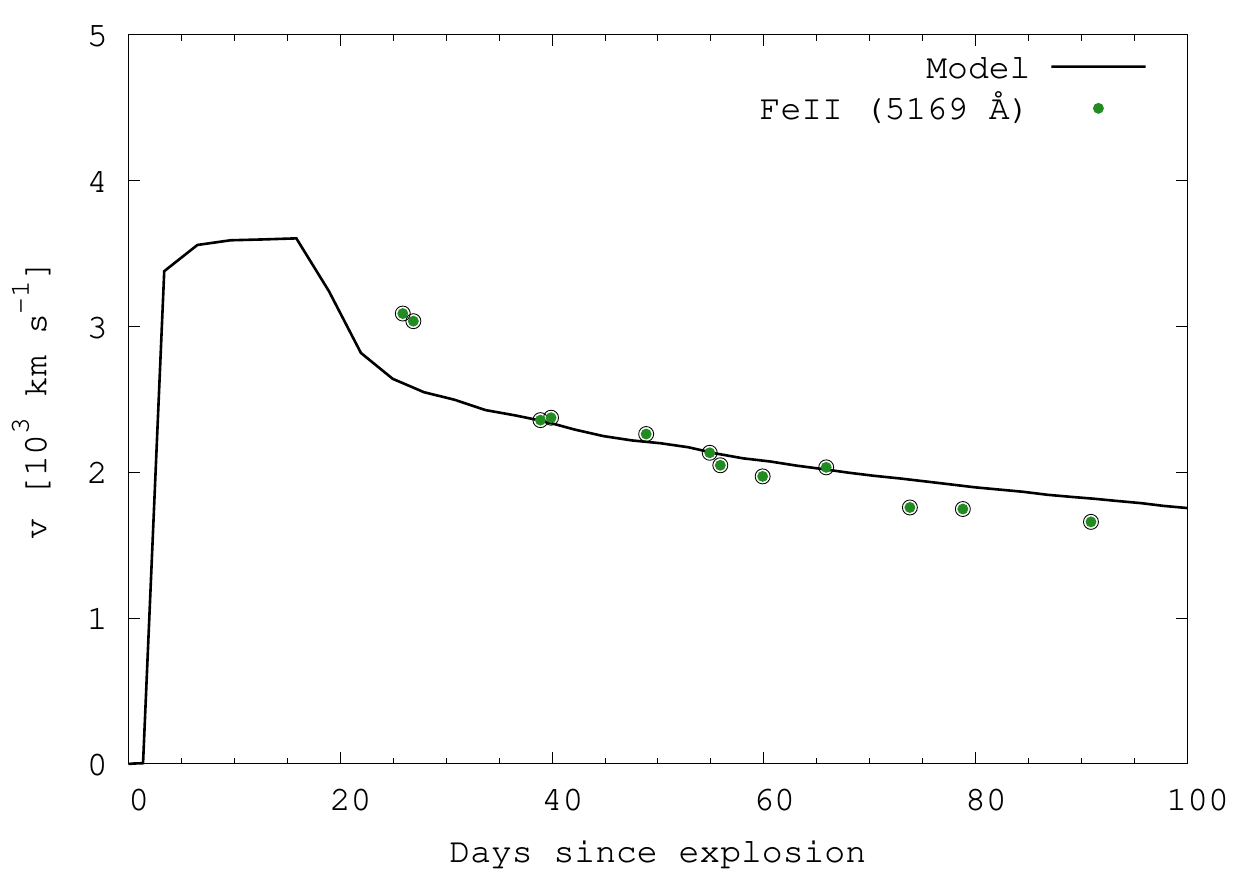}
        \begin{flushleft}
                \caption{Comparison between models and observations for SN 2008bk. (Left) Bolometric light curves. (Right) Evolution of the photospheric velocity. In this case we used a 10\,$M_{\odot}$ rotating progenitor at solar metallicity coming from stellar evolution calculations.}
                \label{fig:08bk_se}
        \end{flushleft}
\end{figure*}

Although we have used double polytropic models as an initial configuration, in order to freely choose the radius of the object, here we present the results for one SN in our sample, SN\,2008bk, using a pre-SN  structure calculated by stellar evolution. We  searched for the available stellar evolutionary models that have parameters (mass and radius) close to the values that we found in our modelling for SN\,2008bk. Figure \ref{fig:08bk_se} shows the comparison between the observed bolometric LC and evolution of the photospheric velocity with our model noticing a good agreement in the plateau, transition, and radioactive tail phases. We used a 10\,$M_{\odot}$ rotating progenitor at solar metallicity from \citet{heger00}, denoted E10. The progenitor experienced little mass loss and retained almost all the mass at the core collapse. The mass and radius of the progenitor at the moment of collapse are 9.23\,$M_{\odot}$ and 550\,$R_{\odot}$. In addition, we assume an explosion energy of 0.15 foe and \Ni\, mass of 0.007\,$M_{\odot}$ to reproduce the observations. We would like to point out the good agreement between the physical parameters used to model SN\,2008bk assuming the double polytropic and the stellar evolution configuration (see Table \ref{table:results}).

\section{Election of our preferred models}
\label{ap:models_election}

\begin{figure*}
\centering
                \includegraphics[width=0.46\textwidth]{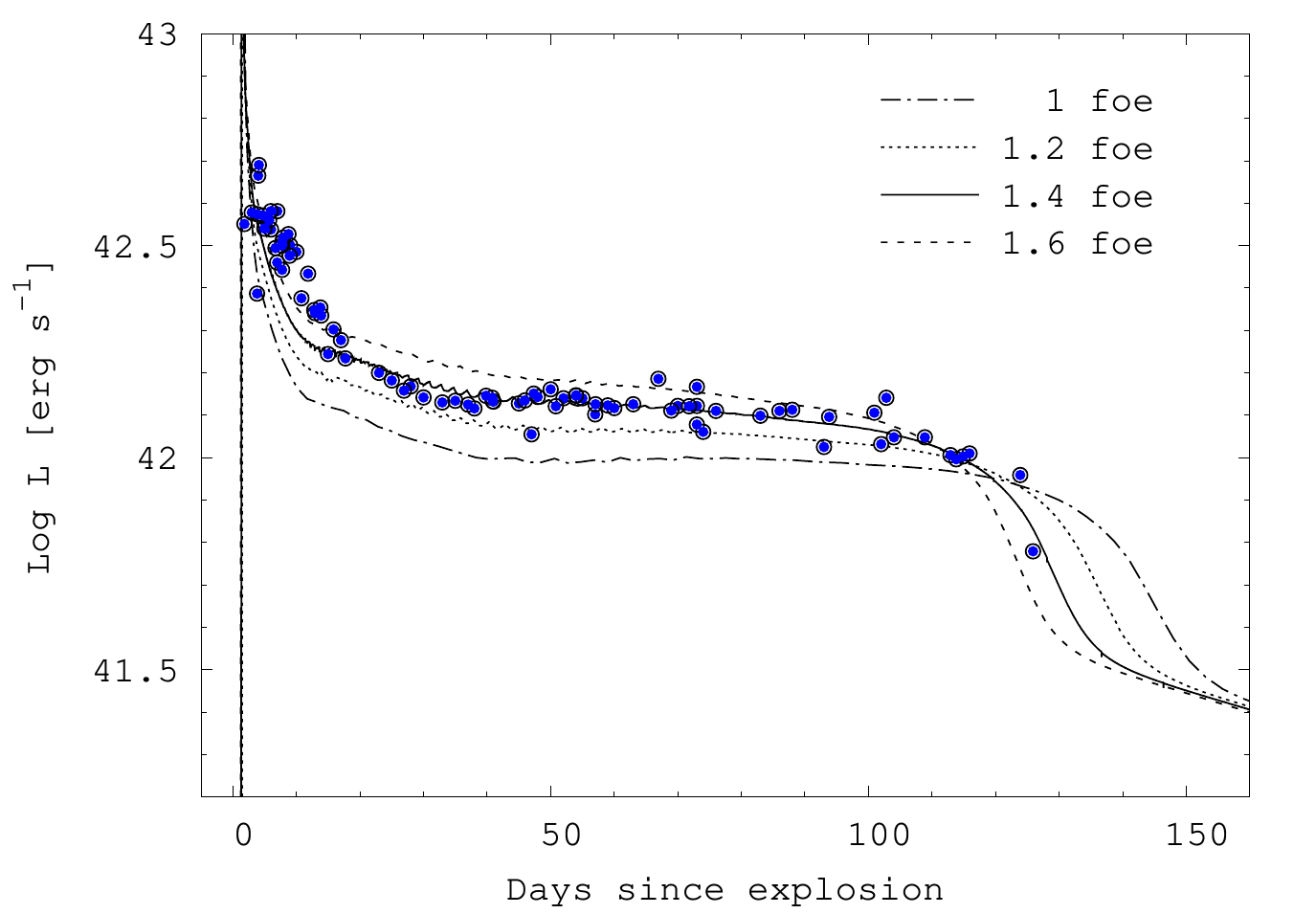} 
                \includegraphics[width=0.46\textwidth]{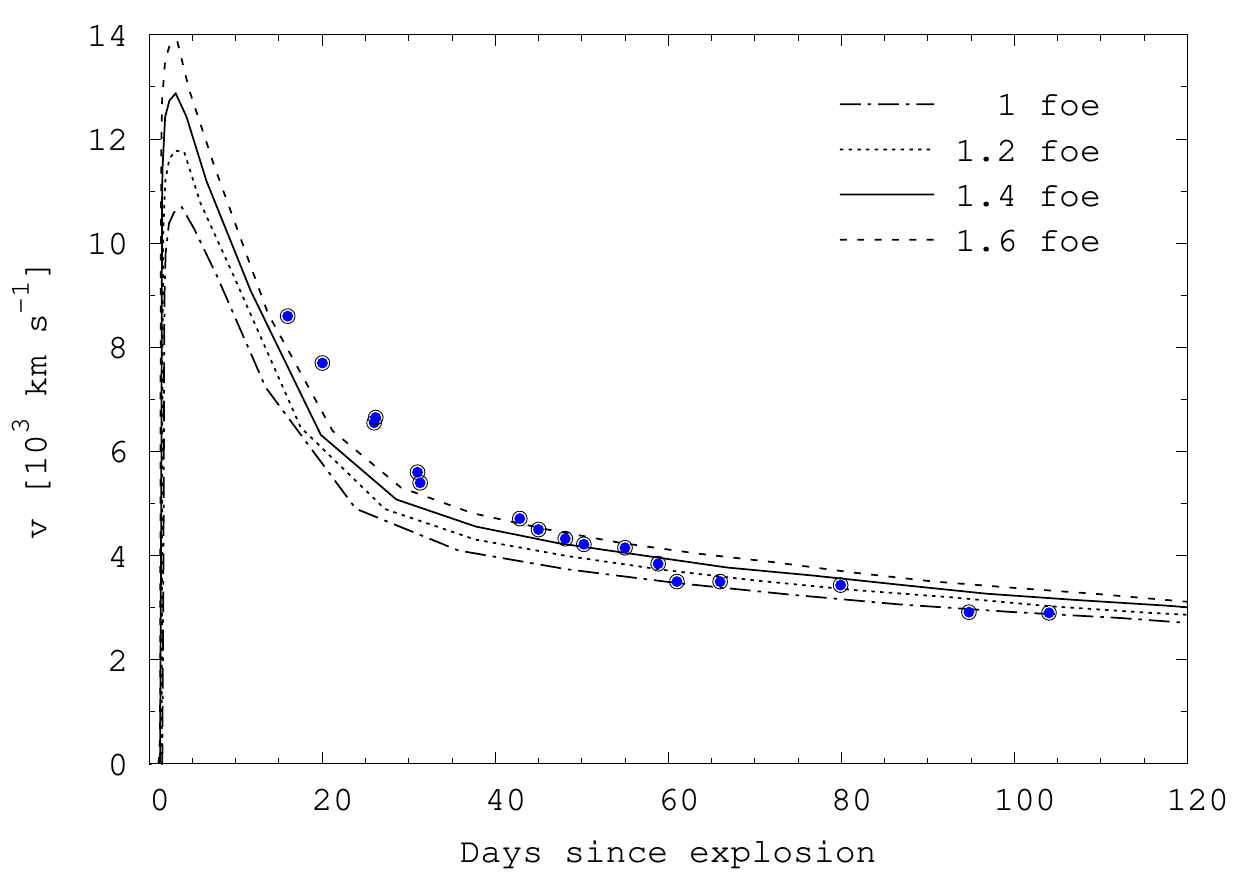} 
        \begin{flushleft}
                \caption{Dependence of bolometric LC (left) and photospheric velocity evolution (right) with the explosion energy for SN 2012aw. The optimal model is shown with a solid line. Models with changes of 0.2 foe or less are still admissible, but beyond these limits, the representation of observations worsens considerably.}
                \label{fig:2012aw_var_energia}
        \end{flushleft}
\end{figure*}

\begin{figure*}
\centering
                \includegraphics[width=0.46\textwidth]{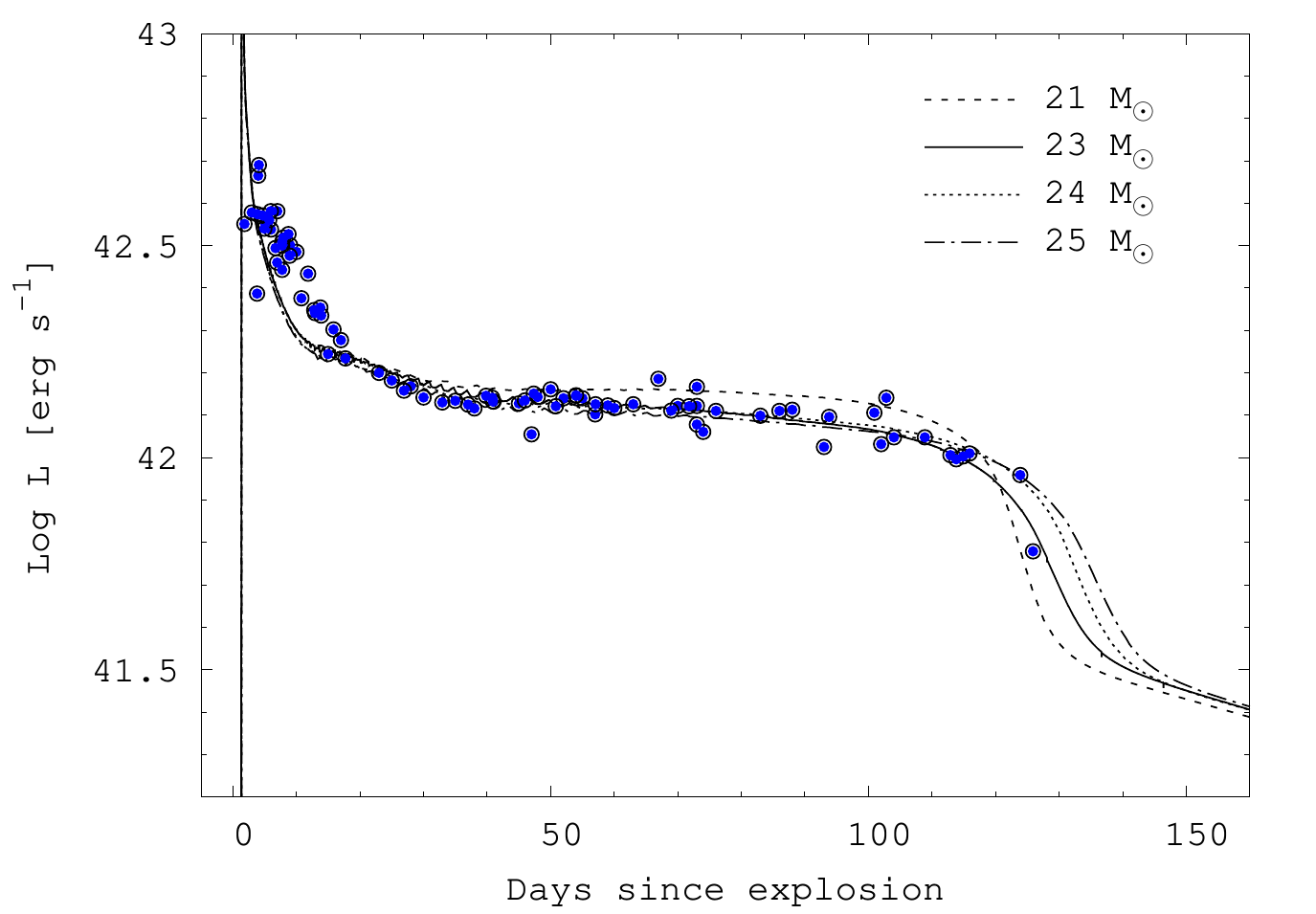} 
                \includegraphics[width=0.46\textwidth]{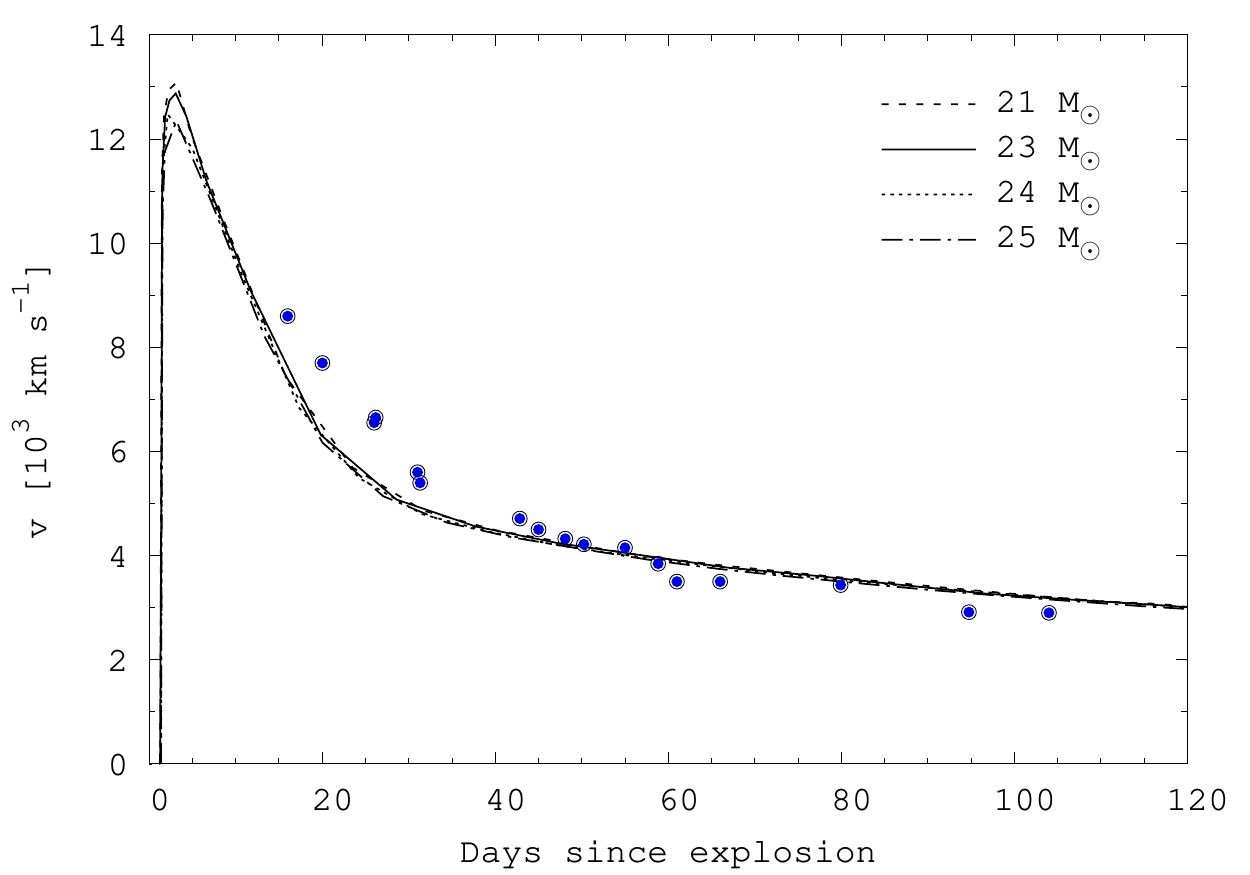} 
        \begin{flushleft}
                \caption{Dependence of bolometric LC (left) and photospheric velocity evolution (right) with progenitor mass for SN 2012aw. The optimal
model is shown with a solid line. A range of masses between 21~and~24\,$M_{\odot}$ still produce acceptable configurations.}
                \label{fig:2012aw_var_masa}
        \end{flushleft}
\end{figure*}

\begin{figure*}
\centering
                \includegraphics[width=0.46\textwidth]{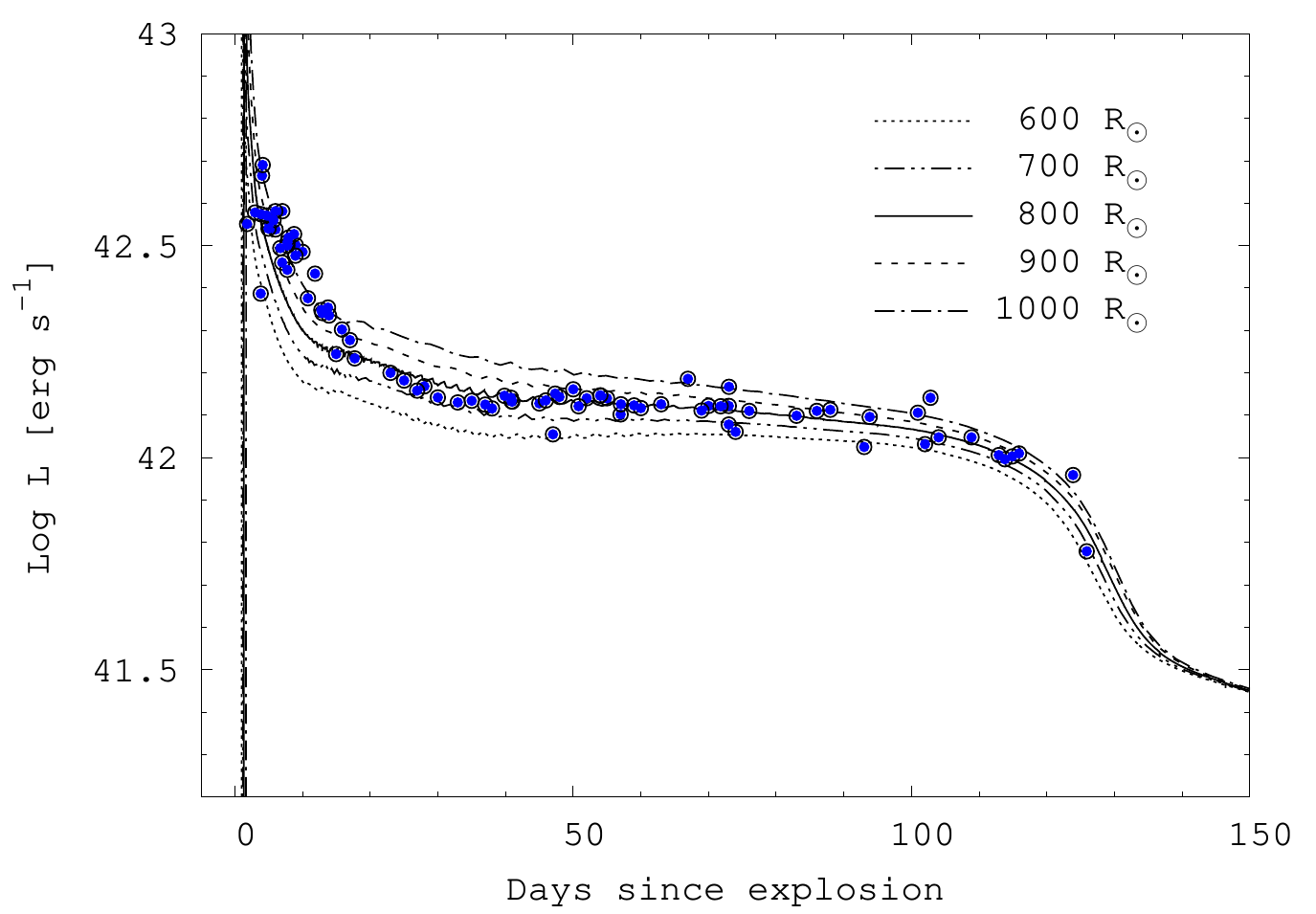} 
                \includegraphics[width=0.46\textwidth]{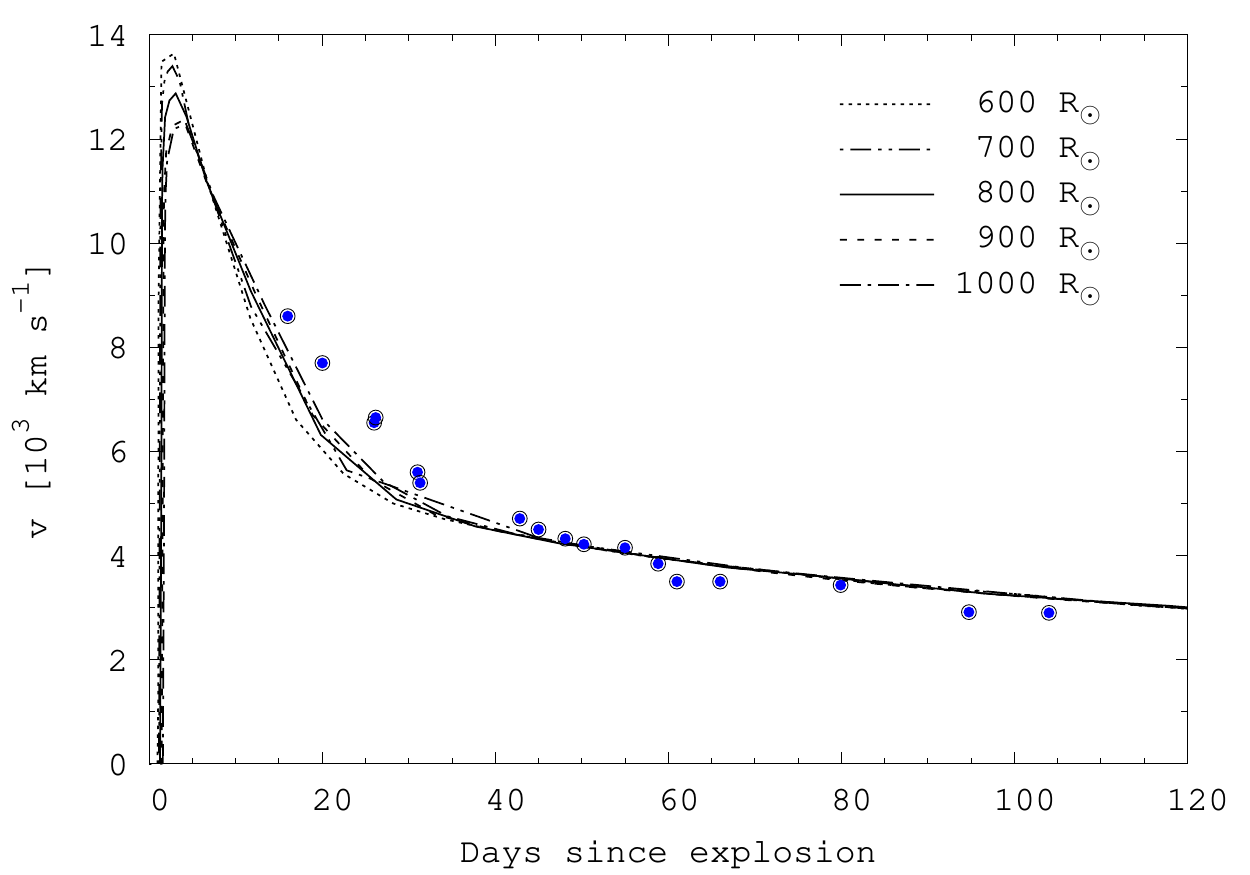} 
        \begin{flushleft}
                \caption{Dependence of bolometric LC (left) and photospheric velocity evolution (right) with pre-explosion radius for SN 2012aw. The optimal
model is shown with a solid line. A range of radius between 700~and~900\,$R_{\odot}$ still produce good representations of observations.}
                \label{fig:2012aw_var_radio}
        \end{flushleft}
\end{figure*}

\begin{figure*}
\centering
                \includegraphics[width=0.46\textwidth]{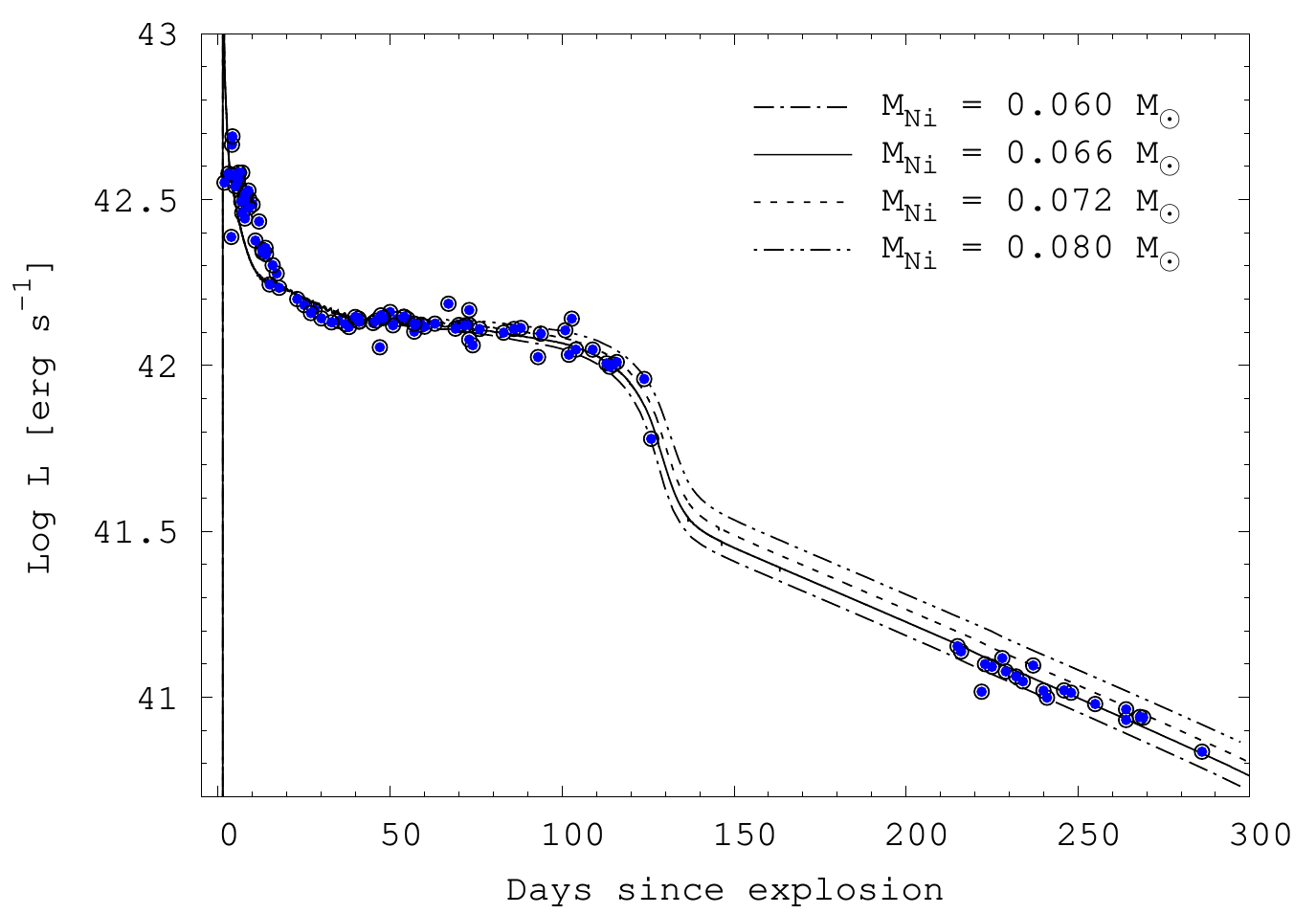} 
                \includegraphics[width=0.46\textwidth]{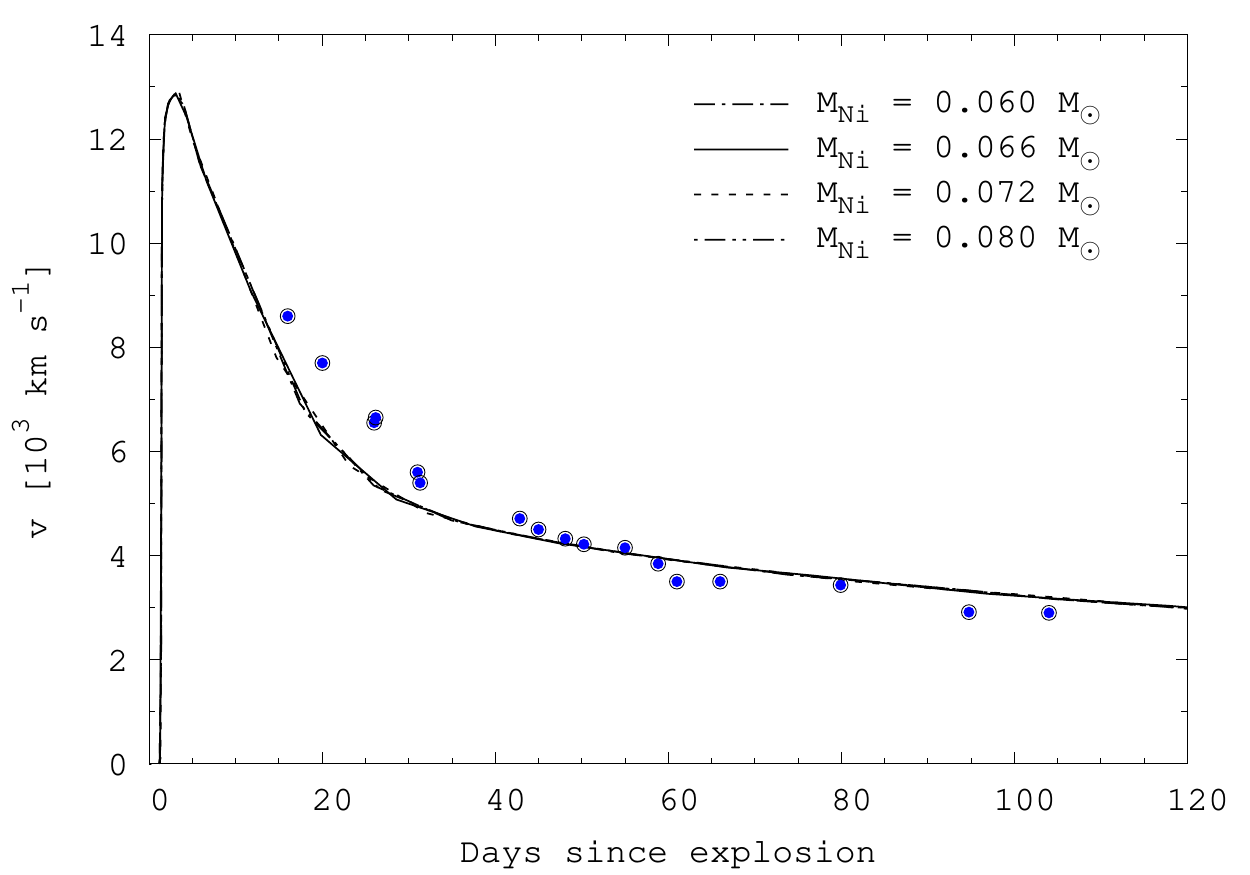} 
        \begin{flushleft}
                \caption{Dependence of bolometric LC (left) and photospheric velocity evolution (right) with $M_{\rm Ni}$. The optimal
model is shown with a solid line. Models with 0.06~--~0.072\,$M_{\odot}$ still produce a suitable representation of data.}
                \label{fig:2012aw_var_niquel}
        \end{flushleft}
\end{figure*}

Here we present an example of how we selected the preferred models and the range of validity for the physical parameters adopted. Although a chi-square minimization would be the appropriate method, due to a lack of knowledge surrounding the errors involved, this method is not used. We note for example that if we were to consider the uncertainty in the distance, the range in the physical parameters would be so big that it would be almost impossible to determine the properties of the progenitor. In Sect. \ref{sec:models} we mentioned that four parameters are necessary to model these quantities: $M_{\rm hydro}$, $R$, $E,$ and $M_{\rm Ni}$, and the degeneracy between the first three of them.

Figures \ref{fig:2012aw_var_energia} - \ref{fig:2012aw_var_niquel} show how LCs and photospheric velocities are modified when one of the parameters is changed, leaving the rest fixed. In this analysis, observational data for SN\,2012aw are presented to show that with small changes in any physical property, models may not represent a good agreement with observations. We present the analysis for SN\,2012aw, but a similar study was done for each SN in our sample. 

In Fig. \ref{fig:2012aw_var_energia} we changed the value for the explosion energy keeping the rest fixed. The plateau length, its luminosity, and photospheric velocity evolution are considerably affected already with changes of $\Delta E =$ 0.2 foe. Increasing that amount of energy to our preferred model ($E$ = 1.4 foe, solid line), the plateau length decreases and the SN becomes more luminous, resulting in a clearly less favourable configuration than our optimal model. This becomes even worse if we keep increasing the energy. When analysing the model with $E =$ 1.2 foe (0.2 foe below the optimal), the behaviour of the LC is opposite to that mentioned before. Hence, a considerably increase in the plateau length and a decrease of plateau luminosity is observed. Again, this produces a poorer representation of observations than our preferred model and the same happens if we decrease the energy even more. In these cases, the photospheric velocity evolution is affected, but an acceptable agreement is still achieved. Therefore, we assume an explosion energy of 1.4 foe, but we consider that those models with changes of 0.2 foe or less are still admissible.

We then performed the same analysis but varying the pre-SN mass ($M_{\rm hydro}$). We can see that the model with $M_{\rm hydro}$ = 21\,$M_{\odot}$, that is to say, 2\,$M_{\odot}$ below our preferred model (solid line in Fig. \ref{fig:2012aw_var_masa}), induces an increase of the plateau luminosity causing a poorer representation of observations. In addition, the plateau length decreases, though differences are small. For the model with $M_{\rm hydro}$ = 24\,$M_{\odot}$, the plateau luminosity does not change significantly while the plateau extends by a few days, again without major differences. When analysing larger variations in the progenitor mass, the models produce a poorer match to the data. In particular, in Fig. \ref{fig:2012aw_var_masa} a model with 25\,$M_{\odot}$ is plotted. In this case the plateau length increases considerably. In the aforementioned cases, photospheric velocities are not significantly affected. With all these considerations, we infer that models with $M_{\rm hydro}$ = 23$^{+1}_{-2}$\,$M_{\odot}$ are still admissible.

In Fig. \ref{fig:2012aw_var_radio} the analysis of the variation of the pre-explosion radius is shown. It is noted that LCs with $\pm$ 100\,$R_{\odot}$ with respect to our optimal model ($R =$ 800\,$R_{\odot}$; solid line) still produce a good representation of the data, but outside that range the agreement becomes poorer. Photospheric velocities are not significantly affected for this analysis. Therefore, we consider that models with changes in the pre-explosion radius less than 100\,$R_{\odot}$ are reliable.

Figure \ref{fig:2012aw_var_niquel} shows the comparison between different models when \Ni\, mass is changed. In this case, it is noted again that LCs with $\pm$ 0.006\,$M_{\odot}$ of \Ni\, with respect to our preferred model (solid line in the figure) still present acceptable configurations. When varying the amount of \Ni\, even more, the bolometric luminosity in the tail of the LC is considerably affected, since in this phase, the bolometric luminosity is a direct measurement of the amount of radioactive nickel. This parameter does not affect the photospheric velocity evolution of the SN. Therefore, with these considerations, we assume that $M_{\rm Ni} = 0.066 \pm 0.006$\,$M_{\odot}$.

This test demonstrates how our optimal models were chosen, showing that small variations in any parameter could produce considerable modifications in the models. This analysis allowed us to  identify the ranges of values within which the models are still acceptable. We feel it necessary to emphasise that the effects of the single-parameter variations do not take into account any potential degeneracy in the solution; we only produce variations along the parameter axis.

\end{appendix}

\end{document}